\pdfoutput=1
\documentclass[acmsmall]{acmart}

\usepackage{bookmark}
\usepackage{algorithm}
\usepackage{tcolorbox}
\usepackage{algpseudocode}
\usepackage{amsmath}
\usepackage{multirow} 
\usepackage[table]{xcolor}
\usepackage[frozencache, cachedir=minted]{minted}
\usepackage{enumitem}
\usepackage{xspace}
\usepackage{listings}
\usepackage{xcolor}
\usepackage[nounderscore]{syntax}
\usepackage{makecell}
\usepackage{siunitx}
\usepackage{graphicx}
\usepackage{subcaption}
\tcbuselibrary{minted}

\definecolor{lightgray}{gray}{0.95}
\definecolor{darkgreen}{rgb}{0,0.6,0}
\definecolor{purple}{rgb}{0.58,0,0.82}

\lstdefinestyle{code}{
  basicstyle=\ttfamily\footnotesize,    % smaller font
  backgroundcolor=\color{lightgray},
  frame=single,
  breaklines=true,
  numbers=none,                          % remove line numbers
  showstringspaces=false,
  aboveskip=4pt,                         % tighter spacing above
  belowskip=4pt,                         % tighter spacing below
  xleftmargin=1em,                       % less left margin
  framexleftmargin=0.5em,
  framexrightmargin=0.5em,
  framextopmargin=0.3em,
  framexbottommargin=0.3em,
  tabsize=2,                             % smaller indent
  captionpos=b,
  commentstyle=\color{gray},
  keywordstyle=\color{blue},
  stringstyle=\color{orange}
}

\newcommand{\tool}{\textsc{WebTestPilot}\xspace}
\newcommand{\pinata}{\textsc{PinATA}\xspace}
\newcommand{\naviqate}{\textsc{NaviQAte}\xspace}
\newcommand{\lavague}{\textsc{LaVague}\xspace}
\graphicspath{{./figures/}}

\setcopyright{cc}
\setcctype{by}
\acmDOI{10.1145/3797115}
\acmYear{2026}
\acmJournal{PACMSE}
\acmVolume{3}
\acmNumber{FSE}
\acmArticle{FSE087}
\acmMonth{7}
\acmSubmissionID{fse26maina-p3243-p}
\received{2025-09-12}
\received[accepted]{2025-12-22}

\begin{document}

\title{WebTestPilot: Agentic End-to-End Web Testing against Natural Language Specification by Inferring Oracles with Symbolized GUI Elements}

\author{Xiwen Teoh}
\orcid{0009-0009-8528-9088}
\affiliation{%
  \institution{National University of Singapore}
  \country{Singapore}
}
\email{xiwen.teoh@u.nus.edu}

\author{Yun Lin}
\authornote{Corresponding author.}
\orcid{0000-0001-8255-0118}
\affiliation{%
  \institution{Shanghai Jiao Tong University}
  \country{China}
}
\email{lin_yun@sjtu.edu.cn}

\author{Duc-Minh Nguyen}
\orcid{0009-0003-2763-6404}
\affiliation{%
  \institution{Shanghai Jiao Tong University}
  \country{China}
}
\email{minh.nguyen@sjtu.edu.cn}

\author{Ruofei Ren}
\orcid{0009-0000-5189-2860}
\affiliation{%
  \institution{Shanghai Jiao Tong University}
  \country{China}
}
\email{renruofei0120@sjtu.edu.cn}

\author{Wenjie Zhang}
\orcid{0000-0002-2669-1837}
\affiliation{%
  \institution{National University of Singapore}
  \country{Singapore}
}
\email{wjzhang@nus.edu.sg}

\author{Jin Song Dong}
\orcid{0000-0002-6512-8326}
\affiliation{%
  \institution{National University of Singapore}
  \country{Singapore}
}
\email{dcsdjs@nus.edu.sg}

\renewcommand{\shortauthors}{Teoh et al.}

\begin{abstract}
Visual language model (VLM) agents show great promise in automating graphical user interface (GUI) testing against requirements in natural language. 
However, the probabilistic nature of language models can have inherent hallucinations.
Therefore, given a detected inconsistency between the requirement and the web application,
it is hard to distinguish whether it stems from the hallucination or a real application bug. 
Addressing this issue presents two core technical challenges: (1)
limited capability and accuracy in deriving implicit test oracles, where the agent must act as its own oracle to implicitly decide if the application's behavior is correct without guidance, and (2) limited reliability due to probabilistic inference, where an LLM’s inconsistent reasoning undermines its trustworthiness as an oracle. 
% Existing LLM-based approaches fail to derive meaningful test oracles from natural language requirements. Some (\lavague, \naviqate) adopt assertion-less strategies, treating any non-crashing navigation as success, while others (\pinata) rely on memory mechanisms that are insufficient for cross-state contextual reasoning. As a result, they fail to detect bugs that depend on relational, temporal, or data dependencies spanning multiple current and prior states.

We introduce \tool, a neurosymbolic LLM-based approach that addresses both challenges through symbolization.
\tool detects and abstracts critical GUI elements of a web application into symbolic variables.
This design improves reliability by constraining assertion generation to operations grounded in explicitly defined symbols, thereby reducing unconstrained or inconsistent reasoning.
At the same time, it improves accuracy by representing application states and their relationships in a structured symbolic form, which increases the likelihood of the agent recognizing data, causal, and temporal dependencies across states. Together, these capabilities enable \tool to generate reliable and accurate test oracles that capture meaningful implicit expectations derived from test requirements.
To advance research in this area, we build a benchmark of bug-injected web apps for evaluating NL-to-E2E testing. 
The results show that WebTestPilot achieves a task completion rate of 99\%, with 96\% precision and 96\% recall in bug detection, outperforming the best baseline (+70 precision, +27 recall). The agent generalizes across diverse natural language inputs (i.e., those containing typos, grammatical errors, redundant sentences, stylistic restyling, or abbreviations) and model scales (3B--72B). In a real-world deployment with a no-code platform, WebTestPilot discovered 8 bugs during development, including data binding, UI, and navigation issues.
% (1) a symbolization layer that models all page states in the test, capturing data, temporal, and causal dependencies as contextual symbols, and (2) a domain-specific language (DSL) that translates natural language requirements into formal pre- and post-condition assertions on these symbols. This design creates stable, explainable test oracles that capture implicit, context-sensitive requirements and enable repeatable validation. To advance research in this area, we build a benchmark of bug-injected web apps for evaluating NL-to-E2E testing. 
% The results show that WebTestPilot achieves a completion rate of 84\%, with 86\% precision and 80\% recall in bug detection, outperforming the best baseline (+36 precision, +54 recall). The agent generalizes across diverse natural language inputs (i.e., those containing typos, grammatical errors, redundant sentences, stylistic restyling, or abbreviations) and model scales (3B--72B). In a real-world deployment with a no-code platform, WebTestPilot discovered 8 bugs during development, including data binding, UI, and navigation issues.
\end{abstract}




\maketitle

\section{Introduction}

The global progressive web application market is projected to reach USD 9.4 billion by 2030 \cite{grandview2024pwa}. As web applications grow in scale and complexity, companies turn to end-to-end (E2E) testing to safeguard reliability for end users, in which testers translate requirements into executable scripts (e.g., Selenium, Playwright, Cypress) that simulate user interactions and verify that applications behave as intended through their end-user interfaces. Without such safeguards, unchecked bugs can escalate into failures that have caused high-profile breakdowns \cite{healthcaregov-fail,queenslandpayroll-fail}.

E2E testing has two main branches. \textit{Exploration-based testing} explores all possible states of a web application to maximize coverage. \textit{Specification-based testing} \cite{robotbt} verifies that the web application behaves consistently with business requirements. Most prior work targets the former, using techniques such as random exploration \cite{Monkey,gremlin.js}, model-based testing \cite{crawljax,atusa}, search-based testing \cite{dig,subweb,feedex,robotest,sapienz,timemachine}, symbolic execution \cite{apollo}, and reinforcement learning \cite{autoblacktest,webexplor,webqt,webrled,unirl,pirltest}. While effective for finding vulnerabilities and corner cases, these methods ignore documentation produced during development, and thus fail to capture meaningful user behaviors. 

In this work, we look into the latter branch. We transform natural language requirements drawn from any source (i.e., UX/UI specifications, product requirements, technical designs, quality assurance plans, or API documents) into executable test actions. Existing approaches both in industry (Cucumber \cite{cucumber}, RSpec \cite{rspec}, Squish \cite{squish}) and academia (GUIPilot \cite{guipilot}, Appflow \cite{appflow}) also look into requirements but rely on rigid input formats (e.g., Sketch files, Gherkin) compatible with their parser. In contrast, we propose a flexible framework that generates tests with verifiable oracles from any natural language excerpt, which (1) directly validates applications against business requirements, (2) speeds up testing for continuous integration and deployment, and (3) reduces maintenance by regenerating tests when requirements change.

Recent advances in large language model (LLM) agents with multimodal reasoning open new possibilities for specification-based GUI testing. According to the State of Software Quality Report \cite{SOSQR-2024} in 2024, over 58\% of respondents use LLM-based tools in automated testing, yet adoption remains limited by capability gaps (44\%) and reliability concerns (30\%). When an agent flags an inconsistency, it is unclear whether the issue stems from the agent itself (hallucination) or the web application (a real bug). Effective automated testing must distinguish between these sources, which give rise to our two key technical challenges:

\noindent\textbf{Limited capability and accuracy in deriving test oracles.} Automated E2E testing requires the agent to act as its own oracle, which is non-trivial \cite{baral2024automating}. An effective test oracle must infer underlying test requirements and translate implicit expectations into concrete assertions. These assertions are only meaningful if they are grounded in data (values reflect prior inputs and computations), causal (state transitions result from the intended actions), and temporal (changes in states referencing the same page over time) dependencies across one or more states. For example, verifying that a ``\textit{product has been added to cart}'' requires not only checking the newly added item, but also ensuring consistency with existing items in the cart (i.e., product types, quantities, and subtotals). Existing works such as \naviqate \cite{naviqate} and \lavague \cite{lavague} lack oracle capability. They translate test requirements directly into actions without generating assertions. Although \pinata \cite{pinata} generates assertions, it relies on a global memory that is (1) precomputed, (2) unstructured, and (3) capacity-limited.

Consider a shopping scenario. On the cart page, \pinata preemptively stores cart items in free-form natural language (e.g., ``Cart contains: Laptop -- \$1200, Mouse -- \$25'') and carries this textual summary forward throughout test execution. This design leads to three limitations:

\begin{enumerate}[leftmargin=*, topsep=2pt]
    \item \textit{Loss of recoverability due to precomputed memory.}
    Only information explicitly recorded at observation time is preserved. If the agent later reaches checkout and intermediate values (e.g., subtotal or applied discounts) were not stored, it cannot reconstruct how the final total was derived. Without these transformation links, it cannot verify whether the total is correct.
    
    \item \textit{Loss of dependencies due to unstructured memory.}
    Historical information accumulates as loosely organized natural language snippets without symbolic identifiers across states. To verify that ``the checkout total equals the sum of item prices minus discount,'' the agent must retrieve fragments from noisy text. Lacking structured references (cart $\to$ shipping $\to$ checkout), it may confuse entries or miss updates, making dependency reasoning brittle.
    
    \item \textit{Loss of changes over time due to capacity-limited memory.}
    As execution lengthens, earlier states are summarized or truncated. For example, if a user applies a 10\% discount and then removes an item, the specification requires the discount to be recalculated. If only the final total is retained, the agent cannot verify whether the recalculation occurred after the removal.
\end{enumerate}

\begin{figure}[t!]
    \centerline{\includegraphics[trim=0 25 0 5, clip, scale=0.27]{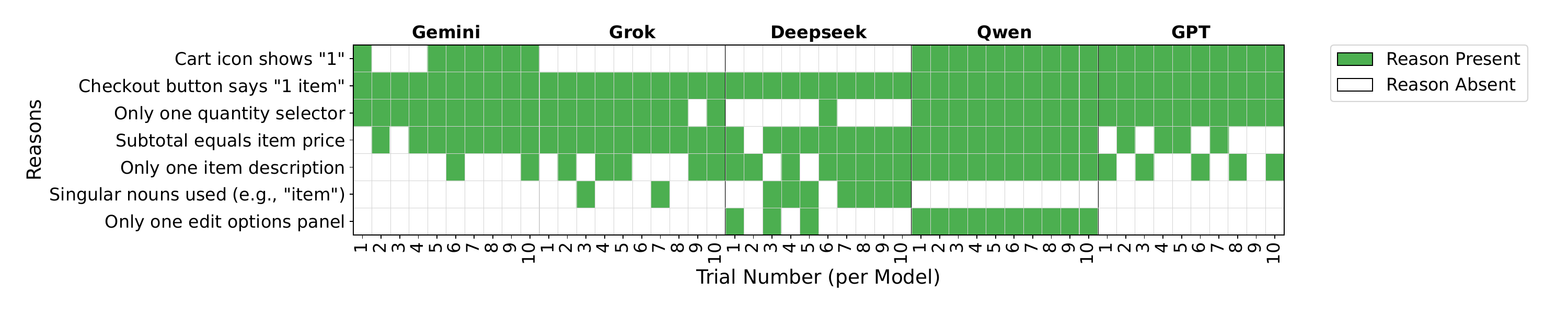}}
    \caption{Inconsistent reasoning by different LLMs with multiple trials in test state verification.}
    \label{fig:inconsistent-models}
    \vspace{-5mm}
\end{figure}

\noindent\textbf{Limited reliability due to probabilistic inference.} By design, LLMs are stochastic: the same prompt can yield different responses even with fixed model settings. When tasked with verifying states during testing, this randomness can lead to inconsistent reasoning. To illustrate, we conducted an empirical study (Figure~\ref{fig:inconsistent-models}): five mainstream LLMs (GPT, Gemini, Grok, Deepseek, and Qwen) were each prompted 10 times with the same page screenshot and the question, ``\textit{Does the shopping cart contain only one item?}'' Across trials, the models produced seven distinct answers, with reasoning varying both across and within models. This variability has visible consequences for multi-step tests. Consider a test with $n$ sequential steps, where each step relies on correct reasoning from the LLM. Even if the probability of a single step being correct ($p$) is high, the stochastic nature of the model means that completing the full trajectory successfully becomes unlikely ($p^n$), as errors compound across steps. Inconsistent outputs in individual steps produce flaky end-to-end verdicts, and interpreting the models’ natural language reasoning adds further manual overhead, breaking the assumption of a stable test oracle in automated testing.

On one hand, effective test oracles must establish data, causal, and temporal dependencies across states to infer implicit expectations that satisfy test requirements. On the other hand, free-form reasoning and verification with stochastic LLMs is inherently unstable. To address these challenges, we propose \tool, a neurosymbolic approach capable of acting as its own accurate and reliable test oracle. \tool uses symbolization to uniformly improve both oracle accuracy and reliability. Building on the success of effective perception models \cite{omniparser} to extract symbols from states, our approach is guided by two key insights: (1) Symbolization improves reliability by converting test oracle generation from a continuous space into a discrete one with finite bounds. By defining explicit symbols, the set of possible assertions is constrained, which reduces uncertainty and guides the agent towards assertions that are semantically valid and consistent. In addition, \tool can perform retrials when generating assertions to reduce hallucinations and maintain stability. and (2) Grounding assertions in symbols improves accuracy. By representing states and their relationships as structured symbols, the agent can explicitly track how UI elements evolve across states. the agent is more likely to recognize data, causal, and temporal dependencies across states. This structured exposure increases the chance that generated assertions capture implicit expectations and faithfully reflect the intended behavior of the application. However, designing this approach involves two technical challenges:

\noindent\textbf{How to link symbols with assertions?} 
After extracting symbols, the agent must compose them into correct, executable assertions. The challenge is designing a domain-specific language (DSL) that balances expressiveness and simplicity: it must be rich enough to capture application behaviors, yet simple enough to avoid hallucination or retraining. We address this by extending an existing programming language, using its familiar syntax and native libraries for data processing, while providing a predefined set of operators over symbols (e.g., relational and compositional predicates).

\noindent\textbf{How to achieve effective and efficient symbolization?} A naive approach would symbolize all visible UI elements and track dependencies for every symbol, leading to combinatorial explosion and the same limitations as global memory in \pinata. We instead propose \emph{page reidentification}. It assigns consistent identifiers to logically equivalent pages (e.g., two or more states pointing to the Cart page) and maintains a structured \emph{Session} history of states. Rather than symbolizing eagerly, symbols are derived \emph{on demand} by retrieving states with the same page identifier and extracting only the relevant elements. It enables focused (fewer symbols) and scalable (more states) reasoning.

Specifically, given a natural language test requirement, \tool decomposes it into $n$ (condition, action, expectation) steps. For each step, \tool translates the condition and expectation into pre- and post-condition assertions. It then applies symbolization to extract relevant UI components as symbols, which are composed via a DSL to construct executable assertions satisfying the specified constraints. To support cross-state reasoning, \tool uses page reidentification to detect revisited pages and maintain a structured history of test states.

We evaluate \tool on a newly constructed benchmark of four bug-injected web applications, comparing its performance against three LLM-based GUI testing baselines (\naviqate~\cite{naviqate}, \lavague~\cite{lavague},~\pinata~\cite{pinata}). Our results show that \tool achieves a test completion rate of 99\%, with 96\% precision and 96\% recall in bug detection, outperforming the strongest baseline by +70 precision and +27 recall. \tool is robust across diverse natural language inputs (i.e., those containing typos, grammatical errors, redundant sentences, stylistic
restyling, or abbreviations) as well as across model scales from 3B to 72B parameters. In a real-world deployment with our collaboration partner, a no-code platform, \tool discovers 8 bugs during development.

In summary, our contributions are as follows:

% Neuralsymbolic approach -> symbols -> guide generation of test scripts
% New neurosymbolic framework for GUI testing -> China Mobile
% Extensive evaluation, source code is available at []

\begin{itemize}[leftmargin=*, topsep=2pt]
\item \textbf{Methodology}: We propose the first neurosymbolic GUI testing approach. The neural component extracts symbols from application states to capture dependencies. The symbolic component constructs assertions over the properties, values, and relations of these symbols, ensuring that implicit expectations in the test requirements are satisfied.

\item \textbf{Implementation}: We present \tool, a framework realizing our approach, which has been successfully adopted by our industry collaborator, China Mobile.

\item \textbf{Experiments}: We build a benchmark of four open-source, real-world web applications with 100 injected bugs. We evaluate \tool against LLM baselines on this benchmark and in real-world settings (industry collaborations and GitHub issues), showing that it outperforms state-of-the-art methods in bug detection.

\end{itemize}

The source code for \tool and the benchmark are available at \url{https://github.com/code-philia/WebTestPilot}.

\section{Motivating Example}

\begin{figure}[t]
    \centerline{\includegraphics[trim=0 10 0 5, clip, scale=0.7]{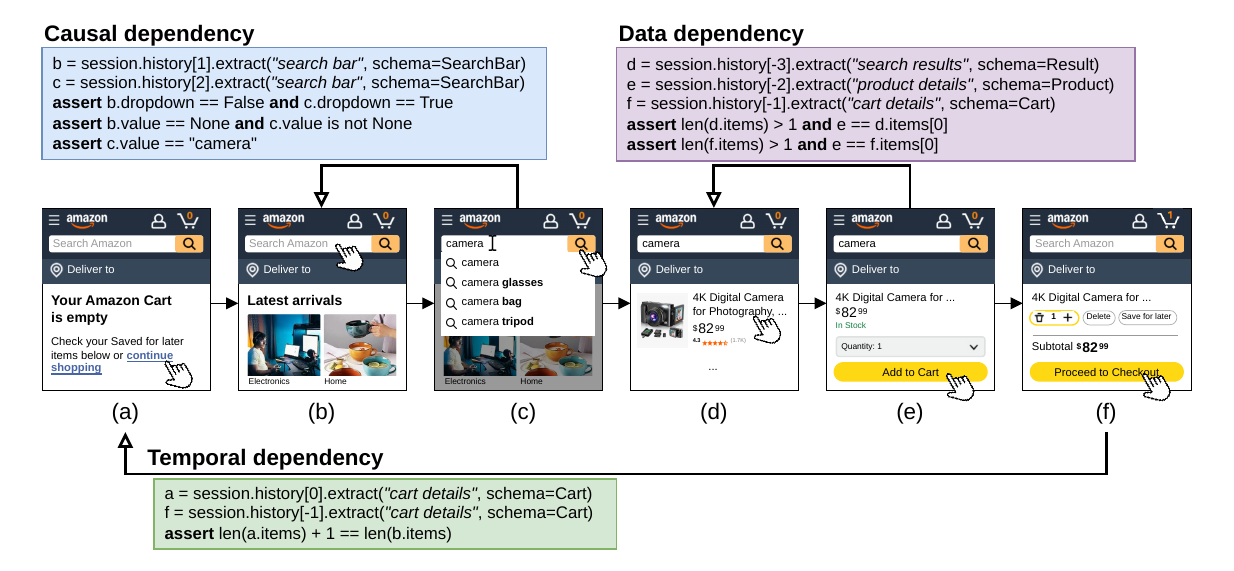}}
    \caption{A test flow depicting search, shopping, and checkout on e-commerce platform \texttt{amazon.com}.}
    \label{fig:motivating-example}
    \vspace{-5mm}
\end{figure}

Figure \ref{fig:motivating-example} shows a test flow on \texttt{amazon.com}, a representative e-commerce scenario. The flow begins on the cart page (State \ref{fig:motivating-example}(a)) with an empty cart. The user clicks “Continue Shopping” to navigate to the homepage (State \ref{fig:motivating-example}(b)), enters the query “camera” in the search bar, triggering a suggestion dropdown (State \ref{fig:motivating-example}(c)), and submits the query to reach a results page (State \ref{fig:motivating-example}(d)). The user then clicks the first product to view its details (State \ref{fig:motivating-example}(e)) and adds it to the cart, completing the test flow.

Applying LLM agents to verify such flows is challenging because meaningful test oracles must reason over dependencies across states, not just the correctness of individual steps. Prior work, such as \naviqate \cite{naviqate} and \lavague \cite{lavague}, considers reaching the end as success without verifying intermediate states. Using the scenario above, if they successfully navigate from State (a) to (f), then the test case passes.
Although \pinata \cite{pinata} maintains a memory to store information and compare it against expected outcomes, its general and unstructured design limits its ability to retrieve task-relevant context for constructing assertions. For example, to verify that the cart subtotal increases exactly by the price of the newly added product, it may fail to recognize that states (a) and (f) correspond to the same logical page, preventing detection of inconsistent incremental changes (i.e., subtotal difference (f) - (a) equals the price of the newly added product). Similarly, to verify that every selected search attribute (e.g., the price range) is preserved across the search results (d), which may include hundreds of products, an unstructured memory may omit a single attribute, leading to incomplete verification.
Many real bugs arise from such inconsistencies. To catch them, it is necessary to reason about the implicit \emph{causal}, \emph{data}, and \emph{temporal} dependencies between states, which are explained below:

\begin{itemize}[leftmargin=*,topsep=2pt]
    \item \textbf{Causal Dependency}: A relation between adjacent states that holds when UI elements in the current state are created, updated, or deleted as a direct effect of executing an action in the previous state. For example, the auto-complete suggestion dropdown and the populated search input in state (c) depends on the typing action in state (b).

    \item \textbf{Data Dependency}: A relation between states that holds when information extracted in one state is propagated to and reused in another, forming a data flow across the execution trace. For example, the product details in (e) depend on the selected item from the search results in (d), and the cart items in (f) depend on the product details in (e).

    \item \textbf{Temporal Dependency}: A relation between states corresponding to the same logical page that holds when a later state must be interpreted relative to an earlier state to detect incremental changes over time. For example, state (f) depends on state (a), both representing the cart page, to determine how the cart contents have evolved after user actions.
\end{itemize}

To enable robust cross-state verification of implicit expectations, \tool supports declarative schemas, which act as symbol templates (or “variables”) representing structured UI data. The schemas are implemented as strongly typed models that can automate parsing, normalization, and validation of extracted content. They define not only the expected data structure, but also type-level constraints (e.g., supported strings), field-level requirements (e.g., required vs. optional), and domain-specific rules (e.g., non-negative prices).

Concretely, consider a test step where after executing the action ``click Add to Cart.'', its corresponding expectation is ``the product is now in the cart.'' To act as a test oracle for this post-condition, \tool first applies symbolization to define relevant symbols (Figure \ref{fig:product-cart-models}).

\noindent
\begin{figure}[H]
\centering
\begin{minipage}[t]{0.48\textwidth}
\scriptsize
\begin{minted}[frame=single]{python}
class Product(BaseModel):
    title: str = Field(...)
    price: float = Field(..., ge=0)
    quantity: Optional[int] = Field(None, gt=0)
\end{minted}
\end{minipage}%
\hfill
\begin{minipage}[t]{0.48\textwidth}
\scriptsize
\begin{minted}[frame=single]{python}
class Cart(BaseModel):
    items: List[Product] = Field(...)

    
\end{minted}
\end{minipage}
\caption{Definition of the \texttt{Product} and \texttt{Cart} symbols, represented as Pydantic schemas.}
\label{fig:product-cart-models}
\vspace{-0.5em}
\end{figure}

\tool then instantiates these schemas with values extracted from the current and prior states. By referencing page reidentification, it recognizes that State (a) and State (f) correspond to the Cart page and learns a high-level overview of its layout (e.g., the cart contains a list of items, each displaying specific information), allowing the Cart symbol to be applied. Similarly, it identifies State (e) as the Product Detail page, where the Product symbol is relevant. By combining this information with the historical actions, the agent establishes the semantic connection of adding a product to the cart through the transitions State (a) $\to$ State (e) $\to$ State (f). See Figure \ref{fig:cart-extraction}.

\begin{figure}[H]
\centering
\scriptsize
\begin{minted}[frame=single]{python}
# State (a): Extract previous cart summary from the initial state
prior = session.history[0].extract("Get cart summary", schema=Cart).items

# State (e): Extract added product from the product details page
added = session.history[-2].extract("Get product detail", schema=Product)

# State (f): Extract current cart summary from the latest state
current = session.history[-1].extract("Get cart summary", schema=Cart).items
\end{minted}
\caption{Extracting the added product from product page and comparing current and prior cart details.}
\label{fig:cart-extraction}
\end{figure}

Using the DSL, the agent constructs a formal assertion on these symbols to verify that the post-action state satisfies both explicit expectations (the product is in the cart) and implicit expectations derived from prior states (the cart contains the same items as before plus the new product, the product type matches the previously viewed item in (e), and the quantity is 1). See Figure \ref{fig:cart-verification}. This allows \tool to detect bugs from implicit and cross-state causal, data, or temporal violations (e.g., missing/duplicate items, wrong quantities or prices, or UI inconsistencies).

\begin{figure}[H]
\centering
\scriptsize
\begin{minted}[frame=single]{python}
# All product attributes (title, quantity, price) match prior and added items
for prod in prior + [added]:
    match = next((p for p in current if p.title == prod.title), None)
    assert match is not None, f"Product {prod.title} missing in current cart"
    assert match.quantity == prod.quantity, f"Quantity mismatch for {prod.title}"
    assert match.price == prod.price, f"Price mismatch for {prod.title}"

# The cart subtotal correctly reflects the addition of the new product.
prior_subtotal = sum(p.price * p.quantity for p in prior)
added_total = added.price * added.quantity
current_subtotal = sum(p.price * p.quantity for p in current)
assert current_subtotal == prior_subtotal + added_total, "Cart subtotal mismatch"
\end{minted}
\caption{Assertion generated by \tool.}
\label{fig:cart-verification}
\end{figure}

\begin{figure}[t]
    \centerline{\includegraphics[trim=0 5 0 5, clip, scale=0.7]{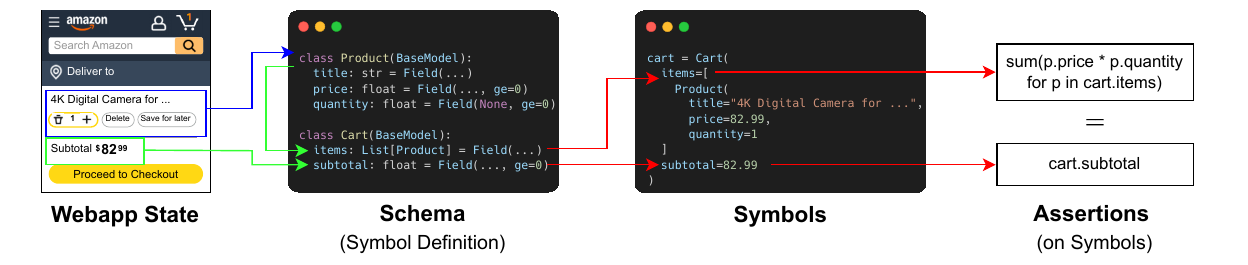}}
    \caption{Example (from motivating scenario): \tool extracts \textit{symbols} via declared \textit{schemas} that correspond to GUI elements for making \textit{assertions} on the application \textit{state}.}
    \label{fig:motivating-example-2}
\end{figure}
\section{Problem Statement}

% \textcolor{red}{TODO: input test requirements, what format? what constraints? what modality? output? ground truth?}

\noindent\textbf{Preliminary.} We model a web application $\mathcal{W}$ as a graph of states $s\in\mathcal{S}$. Each state is defined as a tuple $s=(\text{screenshot},~\text{DOM})$, where $\text{screenshot}$ encodes the visual appearance of the page, and $\text{DOM}$ is a rooted, ordered tree of UI elements $e$, where each element encodes its type (i.e., button, input), relevant attributes (e.g., name, value, enabled/disabled), and child elements. 
A user interacts with $\mathcal{W}$ by executing an action $a = \langle t, e, p \rangle$, where $t$ is the action type (e.g., \texttt{click}, \texttt{type}), $e$ is the target UI element, and $p$ is an optional parameter (e.g., text to enter). The state transition function $\mathcal{T} : \mathcal{S} \times \mathcal{A} \to \mathcal{S}$ maps a state and action to a successor state $s' = \mathcal{T}(s, a)$.
Finally, executing a sequence of actions $A = \langle a_1, a_2, \dots, a_n \rangle$ from an initial state $s_0$ produces an execution trace $\tau=s_0\xrightarrow{a_1}s_1\xrightarrow{a_2}\cdots\xrightarrow{a_n}s_n$ or $s_0 \xrightarrow[\;]{\;A\;} s_n$. 

\noindent\textbf{Objective}. Given a natural language test requirement $D$, an automated tester $T$ parses $D$ into a sequence of steps $\langle\text{step}_1,\dots,\text{step}_n\rangle,~D=\text{step}_1\oplus\cdots\oplus\text{step}_n$, and maps each step to an output $ o_i=T(step_i)$. The details of the input and output are as follows:

\noindent\textbf{Input.} Let the natural language test requirement be a sequence of textual tokens $D=(w_1,\dots,w_m)$. We assume $D$ can be mapped to an ordered sequence of executable steps $\langle \text{step}_1,\dots,\text{step}_n\rangle$, where each $\text{step}_i$ is derived from a contiguous token sequence in $D$. The sequence satisfies:

\begin{enumerate}[leftmargin=*,topsep=2pt]
\item \textit{Disjointness}. The contiguous token sequences corresponding to the steps are pairwise disjoint, i.e., no token in $D$ contributes to more than one $\text{step}_i$.
\item \textit{Monotonicity}. The steps preserve the left-to-right order of $D$: if $\text{step}_i$ is derived from tokens that precede those of $\text{step}_j$, then $i<j$.
\item \textit{Semantic completeness}. Each $\text{step}_i$ encodes: (i) a set of predicates over the input state, (ii) an action to be executed, and (iii) a set of predicates over the output state.
\end{enumerate}

\noindent\textbf{Output.} We define a predicate $p_i$ as a property over application states, $p_i:\mathcal{S}\to\{\top, \bot\}$. We write $s_i \models p_i$ if the state $s_i$ satisfies $p_i$, i.e., $p_i(s_i)=\top$. A bug occurs when $s_i \not\models p_i$ or $p_i(s_i)=\bot$, meaning the state is reported as inconsistent by $T$ with the requirements specified in $\text{step}_i$. During execution, for each parsed step $\text{step}_i$ from $D$, $T$ produces three artifacts $o_i=T(\text{step}_i)=(p_i^\text{pre},~a_i,~p_i^\text{post})$, where:

\begin{enumerate}[leftmargin=*,topsep=2pt]
    \item A predicate $p_i^\text{pre}$ evaluated on the state $s_i^\text{pre}$ before the action.
    \item An action $a_i$ applied on $s_i^\text{pre}$, which transitions $\mathcal{W}$ to a new state $s_i^\text{post}$.
    \item A predicate $p_i^\text{post}$ evaluated on the state $s_i^\text{post}$ after the action.
\end{enumerate}

For $T$ without assertion capability, $p_i^\text{pre}$ and $p_i^\text{post}$ will always evaluate to $\top$.

\begin{figure}[t]
    \centerline{\includegraphics[trim=0 5 0 5, clip, scale=0.6]{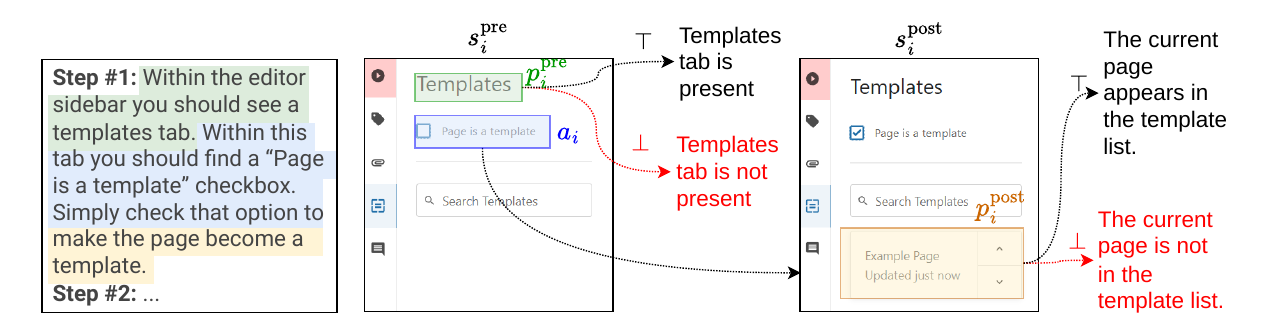}}
    \caption{Visualization of the problem statement’s input and output. $D$ is parsed into steps. The colored overlays highlight regions of interest on the state screenshots: green denotes the portion evaluated by $p_i^\text{pre}$, blue corresponds to the action $a_i$, and yellow denotes the portion evaluated by $p_i^\text{post}$. For each predicate, we illustrate the conditions under which the state is consistent or inconsistent with the requirement $D$.}
    \label{fig:motivating-example-2}
\end{figure}
\section{Approach}

\begin{figure}[t]
    \centerline{\includegraphics[trim=5 10 5 5, clip, scale=0.7]{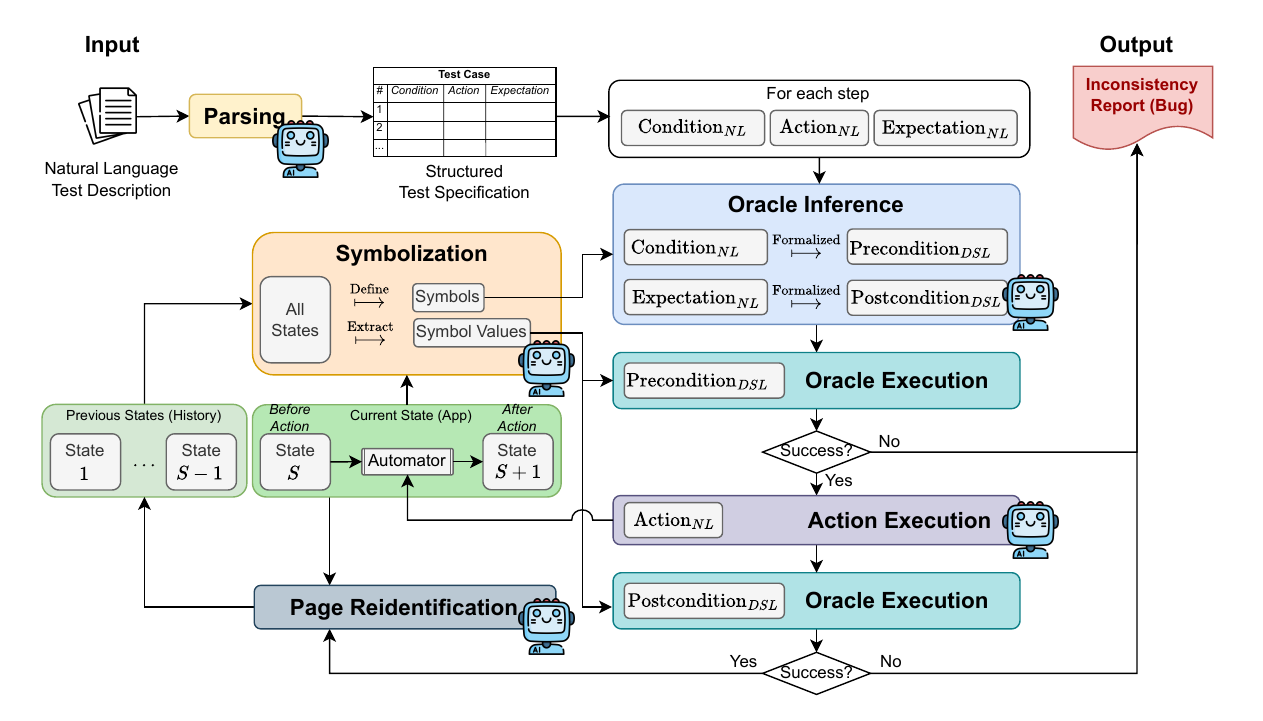}}
    \caption{Overview of \tool. \tool parses a natural language requirement into structured steps (\textbf{Input Parsing}), each specifying a condition, action, and expectation. For each step, it performs \textbf{Oracle Inference} to generate predicate assertions over symbols capturing explicit and implicit requirements. During \textbf{Oracle Execution}, it checks preconditions, executes the action, checks postconditions. Failed assertions trigger retries, and persistent failures are logged as bugs. (\includegraphics[height=1em]{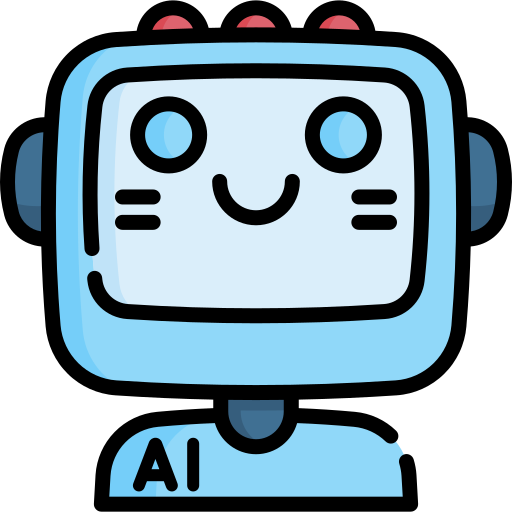}) means the process prompts an LLM.}
    \label{fig:approach}
\end{figure}

\noindent\textbf{Overview} Figure~\ref{fig:approach} shows \tool’s overall approach. Its key novelty is serving as a capable and reliable test oracle, generating predicates $p_i$ that verify implicit expectations from test requirements. This section is organized as follows:

\begin{itemize}[leftmargin=*,topsep=2pt]
\item\textbf{Input Parsing (Section \ref{sec:input_parsing}).} \tool parses a natural language requirement into a structured sequence of steps $\langle\text{step}_1,\dots,\text{step}_n\rangle$, where each step specifies the state before the action ($\text{condition}_\text{NL}$), the action itself ($\text{action}_\text{NL}$), and the state after the action ($\text{expectation}_\text{NL}$).
\item\textbf{Oracle Inference (Section \ref{sec:oracle_inference}).} For each step, \tool analyzes the explicit requirements $\text{condition}_\text{NL}$ and $\text{expectation}_\text{NL}$. It inspects the execution trace $\tau$ to identify temporal, data, and causal dependencies, which it uses to infer implicit requirements. \tool then defines symbols that abstract relevant states and establishes schemas for their expected content. Finally, it uses a DSL to formalize predicate assertions over the symbols from implicit expectations inferred from requirements,
producing $\text{precondition}_\text{DSL}$ and $\text{postcondition}_\text{DSL}$.
\item\textbf{Oracle Execution (Section \ref{sec:oracle_execution}).} With the assertions ready, \tool maps $\text{action}_\text{NL}$ to an executable action on the web application. Before the action, it evaluates $\text{precondition}_\text{DSL}$ to ensure that the current state satisfies the step’s conditions. After the action, it evaluates $\text{postcondition}_\text{DSL}$ to verify the resulting state meets the expected outcome. If any assertion fails, \tool retries the action up to $n$ times. If all retries fail, it reports a bug.
\end{itemize}

\subsection{Input Parsing}
\label{sec:input_parsing}

Input requirements come in many forms, from formal (PRD, user stories) to informal sources (meeting notes, messages, emails). \tool normalizes these into a sequence of steps $\langle \text{step}_1, \dots, \text{step}_n \rangle$, each a 3-tuple $(\text{condition}_\text{NL}, \text{action}_\text{NL}, \text{expectation}_\text{NL})$ specifying when and where an action occurs, how it is performed, and the expected outcome in natural language. To extract this structure, \tool prompts an LLM with the raw input and parses its JSON output. It can be configured to infer and fill-in any missing steps.
\subsection{Oracle Inference}
\label{sec:oracle_inference}

For each step, $\text{step}_i = (\text{condition}_\text{NL}, \text{action}_\text{NL}, \text{expectation}_\text{NL})$, \tool prompts an LLM in two stages. First, the LLM receives the explicit requirements $(\text{condition}_\text{NL}, \text{expectation}_\text{NL})$ and the execution trace in text form $\text{string}(\tau)=[\text{string}(s_0),\dots,\text{string}(s_n)$] (see Section \ref{sec:page_reidentification} for details) to infer implicit requirements by identifying causal, temporal, and data dependencies. Second, the LLM uses $\text{string}(\tau$) together with explicit and implicit requirements to define custom symbols for relevant concepts (e.g., \texttt{Cart}, \texttt{Product}; Section~\ref{sec:state_symbolization}). It then applies the DSL (Section~\ref{sec:dsl}) to generate formal predicate assertions over these symbols. This yields the mapping $\text{condition}_\text{NL} \mapsto \text{precondition}_\text{DSL}$ and $\text{expectation}_\text{NL} \mapsto \text{postcondition}_\text{DSL}$, where preconditions and postconditions are predicate assertions over the starting and ending states $s$ and $s'$ in a step, respectively.

% \begin{figure}[ht]
% \label{fig:dsl}
% \captionsetup{type=figure}
% \caption{BNF syntax of DSL for writing test assertions}
% \scriptsize
% \begin{tabular}{p{0.45\linewidth} p{0.45\linewidth}}
% \toprule
% \texttt{$\Phi$} ::= \textbf{assert} \texttt{Pred} 
%     & \textit{Top-level assertion} \\

% \texttt{Pred} ::= \texttt{Pred and Pred} 
%     \newline\hspace*{2.5em} | \texttt{Pred or Pred}
%     \newline\hspace*{2.5em} | \texttt{not Pred}
%     \newline\hspace*{2.5em} | \texttt{(Pred)}
%     \newline\hspace*{2.5em} | \texttt{Expr comp Expr} 
%     & Boolean logic and comparisons \\

% \texttt{Expr} ::= \texttt{value} 
%     \newline\hspace*{2.5em} | \texttt{var}
%     \newline\hspace*{2.5em} | \texttt{var.attr}
%     \newline\hspace*{2.5em} | \texttt{var.method(args)} 
%     & Values, variables, field/method access \\

% \texttt{comp} ::= \texttt{==}, \texttt{!=}, \texttt{>}, \texttt{>=},
% \texttt{<}, \texttt{<=}, \texttt{in}, \texttt{not in}
%     & Comparison operators \\

% \texttt{args} ::= \texttt{Expr} (',' \texttt{Expr})*
%     & Argument list (comma-separated) \\

% \texttt{var} ::= identifier
%     & Valid variable name in Python \\

% \texttt{attr} ::= \texttt{id} | \texttt{text} | \texttt{children} | $\cdots$
%     & Of \texttt{Session}, \texttt{State}, or \texttt{Element} object \\

% \texttt{method} ::= \texttt{extract()} | \texttt{find()} | $\cdots$
%     & Of \texttt{Session}, \texttt{State}, or \texttt{Element} object \\

% \bottomrule
% \end{tabular}
% \end{figure}

\begin{figure}[ht]
\centering
\begin{minipage}[b]{0.48\linewidth}
\captionsetup{type=figure}
\scriptsize
\begin{tabular}{p{0.45\linewidth} p{0.45\linewidth}}
\toprule
\texttt{$\Phi$} ::= \textbf{assert} \texttt{Pred} 
    & \textit{Top-level assertion} \\

\texttt{Pred} ::= \texttt{Pred and Pred} 
    \newline\hspace*{2.5em} | \texttt{Pred or Pred}
    \newline\hspace*{2.5em} | \texttt{not Pred}
    \newline\hspace*{2.5em} | \texttt{(Pred)}
    \newline\hspace*{2.5em} | \texttt{Expr comp Expr} 
    & Boolean logic and comparisons \\

\texttt{Expr} ::= \texttt{value} 
    \newline\hspace*{2.5em} | \texttt{var}
    \newline\hspace*{2.5em} | \texttt{var.attr}
    \newline\hspace*{2.5em} | \texttt{var.method(args)} 
    & Values, variables, field/method access \\

\texttt{comp} ::= \texttt{==}, \texttt{!=}, \texttt{>}, \texttt{>=},
\texttt{<}, \texttt{<=}, \texttt{in}, \texttt{not in}
    & Comparison operators \\

\texttt{args} ::= \texttt{Expr} (',' \texttt{Expr})*
    & Argument list (comma-separated) \\

\texttt{var} ::= identifier
    & Valid variable name in Python \\

\texttt{attr} ::= \texttt{id} | \texttt{text} | \texttt{children} | $\cdots$
    & Of \texttt{Session}, \texttt{State}, or \texttt{Element} object \\

\texttt{method} ::= \texttt{extract()} | \texttt{find()} | $\cdots$
    & Of \texttt{Session}, \texttt{State}, or \texttt{Element} object \\
\bottomrule
\end{tabular}
\caption{BNF syntax of DSL for writing test assertions}
\label{fig:dsl}
\end{minipage}\hfill
% Put your other figure/table in another minipage here:
\begin{minipage}[b]{0.48\linewidth}
\centering
\includegraphics[trim=0 0 0 0, clip, scale=0.8]{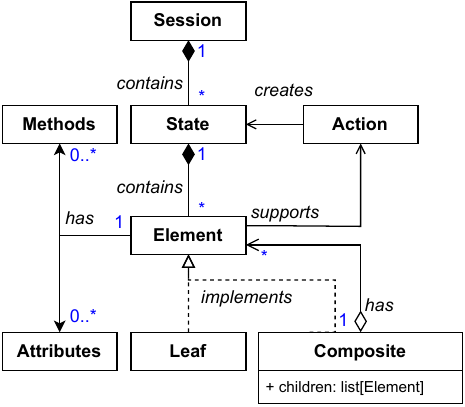}
\caption{Class diagram for built-in symbols.}
\label{fig:class-model}
\end{minipage}
\end{figure}

\subsubsection{State Symbolization}
\label{sec:state_symbolization}

To let \tool reason effectively and identify dependencies, it can abstract domain-specific concepts from any state via custom symbols (e.g., \texttt{Cart} or \texttt{Product}). It defines symbols in Pydantic with type constraints, descriptions, and default values. These symbols can be referenced inside predicate assertions, while their actual values are instantiated at execution time. Predicate assertions are evaluated over the instantiated symbols (see Section~\ref{sec:oracle_execution}).

\subsubsection{Domain Specific Language}
\label{sec:dsl}

To construct and manipulate predicate assertions over symbols, we design a Python-extended domain-specific language (DSL). Its BNF syntax is shown in Figure~\ref{fig:dsl}.

\noindent\textbf{Built-in Symbols.} In addition to custom-defined symbols, the DSL provides a set of general-purpose symbols always available at every step. Figure~\ref{fig:class-model} depicts their class structure. At the top level, \texttt{Session} stores the sequence of states and provides global access to past and current states. Each \texttt{Session} contains multiple \texttt{State} objects, each modeling a specific test step with page metadata and layout information represented as a tree of \texttt{Elements}. The \texttt{State} class offers methods to extract custom symbol values or directly access elements. Table~\ref{tab:methods_and_attributes} lists all the methods and attributes for these symbols.

\noindent\textbf{Expressibility.} By combining custom and built-in symbols, the DSL enables \tool to reason about causal, data, and temporal dependencies. It supports five types of assertions:

\begin{itemize}[leftmargin=*,topsep=2pt]
    \item \textbf{Existence.} Verify the presence or absence of data, e.g., \texttt{state.find("profile") is not None}
    \item \textbf{Relational.} Verify spatial, structural, or logical relationships, e.g., \texttt{state.find("checkout button")[0].ymin > state.find("cart icon")[0].ymax}
    \item \textbf{Temporal.} Ensure events occur in a specific order, e.g., \texttt{all(a.extract(Banner).\\countdown >= b.extract(Banner).countdown for a, b in zip(states, states[1:])}.
    \item \textbf{Causal.} Check cause-effect relations, e.g., \texttt{len(session.history[0].extract(Cart).items) - len(session.history[-1].extract(Cart).items) == 1}.
    \item \textbf{Data Integrity.} Verify extracted or computed data matches expectations, e.g., \texttt{subtotal == sum(item.price for item in cart.items)}.
\end{itemize}

\noindent\textbf{Compositability.} DSL predicates can be single first-order clauses or combinations of multiple clauses connected with logical operators (\texttt{and}, \texttt{or}, \texttt{not}) and grouped with parentheses. Predicates can also span multiple lines of assertions.

\noindent\textbf{Manipulability.} By extending Python, \tool benefits from the LLM’s pre-existing Python knowledge. Its DSL supports Python Standard Library functionality for functional programming (e.g., \texttt{itertools}, \texttt{functools}, \texttt{operator}), text processing (\texttt{re}), built-in functions (\texttt{all()}, \texttt{any()}, \texttt{filter()}, \texttt{map()}, \texttt{len()}, \texttt{min()}, \texttt{max()}, etc.), and data types (\texttt{datetime}, \texttt{enum}). \tool can also control execution with conditional statements and loops.

\begin{table}[ht]
\centering
\caption{Built-in DSL symbols, their attributes and methods}
\label{tab:methods_and_attributes}
\scriptsize
\renewcommand{\arraystretch}{1.05}
\setlength{\tabcolsep}{6pt} % tighten column spacing
\begin{tabular}{lllp{5.4cm}}
\toprule
\textbf{Type} & \textbf{Methods / Attributes} & \textbf{Return Type} & \textbf{Description} \\
\midrule
Session & history & list[State] & Chronological list of all states. \\
       & state & State & Current browser page. \\
\midrule
State   & page\_id & string & Logical page identifier shared across states. \\
       & elements & set[Element] & All state elements (flattened). \\
       & find(description: str, top\_k: int) & list[Element] & Top K elements matching description (may be empty). \\
       & extract(instruction: str, schema: BaseModel) & BaseModel & Extracts a schema-conforming symbol from the state. \\
\midrule
Element & xmin, ymin, xmax, ymax & int & Bounding box coordinates. \\
       & parent & Element & Parent element. \\
       & children & list[Element] & Child elements. \\
       & extract(instruction: str, schema: BaseModel) & BaseModel & Extracts a schema-conforming symbol from the element. \\
\bottomrule
\end{tabular}
\end{table}
\subsection{Oracle Execution}
\label{sec:oracle_execution}

Once \tool infers the $\text{Precondition}_\text{DSL}$ and $\text{Postcondition}_\text{DSL}$ predicate assertions, it executes the step in three stages. First, it executes $\text{Precondition}_\text{DSL}$. Then, it maps $\text{Action}_\text{NL}$ to an executable action $a = \langle t, e, p \rangle$ (see Section~\ref{sec:action-execution}) and executes it on the current state, producing $s \xrightarrow{a} s'$. Finally, it executes $\text{Postcondition}_\text{DSL}$. Formally, a $\text{step}_i$ is successful iff:

$$
(s \models \text{Precondition}_\text{DSL} \wedge s \xrightarrow{a} s') \implies s' \models \text{Postcondition}_\text{DSL}.
$$

\noindent A test case is successful if every $\text{step}_i$ in the sequence is successful. A bug occurs whenever any predicate assertion $p$ at any $\text{step}_i$ fails, i.e., $p(s)=\top,~s\not\models p$.

\subsubsection{Action Execution} 
\label{sec:action-execution}
Set-of-Mark prompting (e.g., OmniParser \cite{omniparser}) is stable but costly and sensitive to noise, such as when the page contains too many elements, while GUI grounding models (e.g., \cite{uground,uivenus}) are fast but unstable, especially in out-of-distribution settings such as unseen websites. Inspired by ScreenSeekeR \cite{screenseeker}, \tool combines these approaches to achieve a balance of stability and efficiency. Given $\text{Action}_\text{NL}$ and the full page screenshot in the current state $s$, \tool uses a GUI grounding model (UI-Venus-7B \cite{uivenus}) to predict coarse target coordinates $(x, y)$, and then apply Set-of-Mark prompting to annotate all interactable elements on the screenshot with bounding boxes and IDs by analyzing the tree of UI elements in $s$. A square crop centered at $(x, y)$, together with $\text{Action}_\text{NL}$, is fed to an LLM, which outputs precise executable actions $\langle t, e, p \rangle$, where $t \in \{\text{click}, \text{type}, \text{press}, \text{scroll}, \text{wait}\}$, $e$ is the element ID, and $p$ are action parameters. This approach mitigates GUI grounding instability by using it for predicting coarse approximate locations, while reducing cost and improving effectiveness by focusing the LLM on a cropped screenshot of the target region.

% \label{sec:action_execution}
% \tool prompts an LLM with $\text{Action}_\text{NL}$ and a state screenshot to infer an action type $t \in \{\text{click}, \text{type}, \text{drag}, \text{scroll}, \text{wait}\}$, a target element description, and parameters $p$. A GUI grounding model (UGround \cite{uground}) maps the description to coordinates $(x, y)$, which are resolved to an element $e$ in the state, forming the executable action tuple $\langle t, e, p \rangle$.

\subsubsection{Page Reidentification}
\label{sec:page_reidentification}
To store a new state $s'$ in $\tau$, \tool assigns it a \text{page\_id} so that later states can be recognized as referring to the same logical page (e.g., returning to the shopping cart page). 
\tool first selects $s''$ from $\tau$ with the smallest DOM tree edit distance to $s'$, then it provides screenshots of $s'$ and $s''$ to an LLM to decide if they belong to the same page. If so, $s'.\text{page\_id} = s''.\text{page\_id}$; otherwise, a new $\text{page\_id}$ is assigned. \tool also generates a textual representation $\text{string}(s') = (\text{page\_id}, \text{summary}, \text{layout})$ and appends $s'$ to $\tau$.

\subsubsection{Retry on Assertion Failure}
When a predicate assertion fails, \tool can regenerate it and retry up to $n$ times to reduce the possibility of LLM hallucination. Alternatively, it can generate $n$ candidate predicates upfront and resolve the outcome via majority voting. As long as the assertion holds, the reliability of test results scales with $n$. In this work, we use $n=1$.
\section{Experiments}

We design our experiments to answer the following research questions:

\begin{itemize}[leftmargin=*,topsep=2pt]
    \item \textbf{RQ1 (Test Flow Completion):} How effectively does \tool generate test trajectories that align with human-authored test scripts, compared to baseline GUI testing agents?
    \item \textbf{RQ2 (Bug Detection):} How effective is \tool at detecting visual and functional faults during GUI testing, relative to existing agent-based baselines?
    \item \textbf{RQ3 (Robustness Evaluation):} To what extent can \tool generate correct test cases when provided with requirements expressed in a varied, unstructured, or freely written natural language, without relying on a fixed input format?
    \item \textbf{RQ4 (Model Comparison)} How much does \tool’s overall performance depend on the capabilities of the underlying language model, and to what extent can its refinement mechanisms compensate when using a lightweight and cost-effective LLM?
\end{itemize}

\subsection{Benchmark Construction}

To the best of our knowledge, existing automated UI testing benchmarks primarily target Android mobile applications \cite{AUITestAgent,guitestingarena,themis,guipilot,droiddefects}. While there are datasets for web navigation \cite{formfactory}, UI understanding, and test scripts \cite{e2egit} they do not focus on E2E bug detection. To address this gap, we construct a benchmark of live web applications for E2E test execution and bug detection. The construction proceeds as follows. First, we identify a set of candidate web applications $\{\mathcal{W}_1,\dots,\mathcal{W}_n\}$ that satisfy our selection criteria. Next, for each application $\mathcal{W}_i$, we curate a set of natural-language test requirements $\{D_{i1},\dots,D_{im}\}$ by referring to the application's user documentation and extracting its key functional features. For each requirement $D_{ij}$, we define an executable test script $\mathcal{A}_{ij}=\langle \alpha_{ij}^1,\dots,\alpha_{ij}^k\rangle$ consisting of test assertions, where each assertion $\alpha_{ij}^k:\mathcal{S}\to \{\top, \bot\}$ is a predicate over system states. Each assertion corresponds to a single expected test step to be parsed from $D_{ij}$ by the evaluated automated tester $T$, and serves as a ground truth evaluation oracle for assessing the correctness of its execution trace $\tau$. To evaluate bug detection, we additionally inject a bug for each $D_{ij}$ in the form $\text{bug}_{ij}:\mathcal{S}\to\mathcal{S}$. When applied to a state $s_t$, it either leaves the state unchanged if the bug should not trigger, or produces a modified state $s_t'$ with buggy behavior. In summary, each benchmark sample is represented as a tuple $(\mathcal{W},~D,~\mathcal{A},~\text{bug})$.

\subsubsection{Web Applications}
We search GitHub for open-source web applications and select those based on five criteria: (1) popularity, with $\ge$5,000 stars; (2) active development, with $>$50 contributors and $>$1,000 commits, and a commit in the past month; (3) maturity, publicly available for $>$5 years; (4) practical relevance, indicated by active deployment, recognizable domain or organization, commercial support, or adoption by well-known entities; and (5) user-facing documentation describing core features. We select the following four web applications:

\begin{itemize}[leftmargin=*,topsep=2pt]
    \item \textbf{BookStack} \cite{bookstack}: A hierarchical documentation management platform with rich text editing.
    
    \item \textbf{Indico} \cite{indico}: An event manager for conferences, meetings, and lectures.
    
    \item \textbf{InvoiceNinja} \cite{invoiceninja}: A business-oriented invoicing platform with multi-step workflows.
    
    \item \textbf{PrestaShop} \cite{Prestashop}: A full-stack e-commerce platform with store management feature.
\end{itemize}

We package the applications into reproducible Docker Compose environments.

\begin{table}[!t]
\caption{Overview of the benchmark and its injected bugs.}
\label{table:benchmark}
\centering
\scriptsize
\renewcommand{\arraystretch}{1.4}
\setlength{\tabcolsep}{10pt}
\begin{tabular}{l l l p{6cm}}
\toprule
\textbf{Web Application} & \textbf{Test Cases} & \textbf{Lines of Code} & \textbf{Example Bug} (\textit{from GitHub Issues})\\
\midrule
BookStack~\cite{bookstack} & 27 & 214,819 & No error message shown when user does not have permission to delete attachment (\href{https://github.com/BookStackApp/BookStack/issues/5323}{bookstack/\#5323}).\\
Indico~\cite{indico} & 25 & 573,316 & "Send" button is missing from request recording in lectures (\href{https://github.com/indico/indico/issues/239}{indico/\#239}).\\
InvoiceNinja~\cite{invoiceninja} & 25 & 1,513,289 &  Generating a PDF statement for a client shows the wrong client name and address (\href{https://github.com/invoiceninja/invoiceninja/issues/10351}{invoiceninja/\#10351}).\\
PrestaShop~\cite{Prestashop} & 23 & 2,234,514 & Clicking a product in "All Stores" send you to the "Order" page not the "Edit Product" page (\href{https://github.com/PrestaShop/PrestaShop/issues/39044}{prestashop/\#39044}).\\
\bottomrule
\end{tabular}
\end{table}

\subsubsection{Natural Language Test Requirements} For each application $\mathcal{W}_i$, we construct $\{D_{i1},\dots,D_{im}\}$. We start by adapting verbatim extracts from user documentation, which typically provides how-to guides for key features, and write scripts that follow the “happy paths” (intended successful usage scenarios). We then extend this initial set by extrapolating additional requirements based on the Create, Read, Update, Delete (CRUD) paradigm. For example, if the documentation describes a book management feature, we create test flows for adding, viewing, editing, and deleting book entries. Throughout this process, we follow ISTQB Certified Tester Foundation Level (CTFL) v4.0 guidelines. This yields 100 test requirements. See \autoref{table:benchmark} for more details.

\subsubsection{Test Scripts} For each test requirement $D_{ij}$, the corresponding test script $\mathcal{A}_{ij}$ consists of sequential test assertions $\alpha_{ij}^t$. During testing, at step $t$, if $T$ proposes an action $a_t$ that transitions the system from state $s_t \xrightarrow{a_t} s_{t+1}$, we evaluate $\alpha_{ij}^t(s_{t+1})$: $\top$ if the resulting state is expected state, and $\bot$ otherwise. This is applied sequentially over the entire execution trace $\tau$. We implement test scripts using Playwright. Figure \ref{fig:yaml-example} shows an example test assertion.

\begin{figure}[ht]
\centering
\begin{minipage}{0.95\linewidth}
\begin{lstlisting}[
  basicstyle=\ttfamily\footnotesize,
  breaklines=true,
  showstringspaces=false,
  frame=single,
  aboveskip=0pt,
  belowskip=0pt
]
action:      Click 'Books' link in navigation
expectation: Books listing page with title 'Books' appears
assertion:   expect(page.get_by_role('heading', name='Books')).to_be_visible()
\end{lstlisting}
\end{minipage}
\caption{An example test assertion for a test step.}
\label{fig:yaml-example}
\end{figure}
\vspace{-1em}

\subsubsection{Injected Bugs} We design a single artificial bug $\text{bug}_{ij}:\mathcal{S}\to\mathcal{S}$ for each test requirement $D_{ij}$. These bugs induce incorrect behaviors while ensuring stable and reproducible experiments by locking application versions. To ensure realism, we examine closed GitHub issues labeled "Bug" from each application repository. From a total of 2,043 issues, we randomly sample 10\%. We perform open coding on the titles and descriptions of the sampled issues to identify meaningful labels, and then conduct a thematic analysis to group these labels into broader bug categories. Two co-authors independently perform the analysis, with a third resolving any disagreements. We exclude crash bugs and purely cosmetic bugs (e.g., layout or positioning issues) that do not affect functionality, as prior work has already addressed them. Based on our analysis, we focus on four categories:

\begin{itemize}[leftmargin=*]
    \item \textbf{Missing UI elements}: Required interface components are absent, breaking feature functionality. For example, in \href{https://github.com/PrestaShop/PrestaShop/issues/22170}{prestashop/\#22170}, the "Configure" button is missing for newly installed modules.
    
    \item \textbf{Data inconsistency}: Information shown to the user does not match expected values. For example, in \href{https://github.com/indico/indico/issues/5197}{indico/\#5197}, the category search results include items that were previously deleted.
    
    \item \textbf{No-op actions}: User actions fail silently or have no effect. For example, in \href{https://github.com/invoiceninja/invoiceninja/issues/11188}{invoiceninja/\#11188}, the filter button in "Customer > Documents" does not sort or filter and always shows the full list.
    
    \item \textbf{Navigation failures}: Pages fail to transition correctly. For example, in \href{https://github.com/PrestaShop/PrestaShop/issues/14796}{prestashop/\#14796}, a logged-in user selecting any option in the back-office menu is redirected to the login page.
\end{itemize}

Our categorization aligns with prior studies on Android applications \cite{guitestingarena}. Following these categories, we manually study the source code of each benchmarked web application and implement the bug in JavaScript. The bug function $\text{bug}_{ij}$ is invoked at every state transition during testing. It will automatically modify the system state according to its behavior.

\subsection{RQ1: Test Flow Completion}
\label{sec:rq1_test_case_generation}

In this section, we evaluate \tool's ability to complete test steps on web applications by parsing natural language test requirements. 
% We compare \tool against baseline agents using quantitative metrics to assess its effectiveness and identify strengths, and perform a qualitative analysis of representative failures to better understand its limitations.

\subsubsection{Baselines} We select three baseline agents for GUI testing, described in detail below.

\begin{itemize}[leftmargin=*,topsep=2pt]
    \item \textbf{\lavague}: The most popular open-source, community-supported multi-agent approach. \lavague follows a two-stage architecture: the \textit{World Model} interprets the user’s objective in the context of the current webpage state to produce the next high-level instruction, while the \textit{Action Engine} translates this instruction into executable automation code. It utilizes both the HTML DOM and a visual screenshot of the page to generate DOM-level actions. \lavague focuses solely on test step completion, without verification or assertion.
    
    \item \textbf{\naviqate}: The first single-agent approach guided by functional descriptions. \naviqate operates through a three-step process:
(1) \textit{Action Planning} uses retrieval-augmented generation (RAG) to identify relevant prior tasks that guide planning;
(2) \textit{Choice Extraction} collects actionable elements from the webpage, ranks them based on relevance to the current step, and annotates their functionality;
(3) \textit{Decision Making} prompts the LLM to select an action using an annotated screenshot.
Like \lavague, \naviqate focuses only on test step completion.

    \item \textbf{\pinata}: The state-of-the-art (SOTA) multi-agent approach that separates planning, execution, and verification into three agents: the \textit{Orchestrator}, \textit{Actor}, and \textit{Assertor}. The orchestrator manages the test flow, instructing the actor to perform UI actions and the assertor to verify outcomes. The actor grounds actions using page screenshots and executes them via code actions, while the assertor checks expected results through visual analysis. All agents share a long-term memory and operate solely on the application's observable state.
\end{itemize}

\subsubsection{Evaluation Metrics} 
\label{sec:rq1_metrics}

Let $\tau = s_0 \xrightarrow{a_1} s_1 \xrightarrow{a_2} \cdots \xrightarrow{a_n} s_n$ denote the execution trace produced by the automated tester $T$ (either \tool or a baseline) for a given test requirement $D_{ij}$. Here, $s_k$ is the system state after step $k$, and $\alpha_{ij}^k \in \mathcal{A}_{ij}$ is the assertion for that step in the test script.  We evaluate the effectiveness of $T$ in completing the test using two metrics:  

\begin{itemize}[leftmargin=*,topsep=2pt]
    \item \textbf{Task Completion (TC)}: A test is considered complete if and only if all state transitions in the execution trace satisfy their corresponding test assertions. Formally: 

\[
\mathsf{TC}_{ij} =
\begin{cases}
1 & \text{if } s_k \models \alpha_{ij}^k, \ \forall k = 1,\dots,|\mathcal{A}_{ij}|\\
0 & \text{otherwise}
\end{cases}
\]

    \item \textbf{Correct Trace (CT)}: Measures the fraction of the test script correctly executed before the first assertion failure (prefix). It quantifies how far $T$ progresses along the test. Formally:
    
\[
\mathsf{CT}_{ij} =
\frac{
    \max \Bigl\{ k \in \{1, \dots, |\mathcal{A}_{ij}|\} \;\Big|\; s_\ell \models \alpha_{ij}^\ell \text{ for all } \ell = 1, \dots, k \Bigr\}
}{|\mathcal{A}_{ij}|}
\]

\end{itemize}

\subsubsection{Experiment Setup}

For each test requirement $D_{ij}$ in our benchmark, we let $T$ parse it into a sequence of steps, where each step is a tuple $\text{step}_t = (\text{condition}_\text{NL}, \text{action}_\text{NL}, \text{expectation}_\text{NL})$. At step $t$, $T$ proposes an action $a_t$ corresponding to $\text{action}_\text{NL}$, transitioning the system from $s_t \xrightarrow[]{a_t} s*{t+1}$. The evaluation environment automatically checks $s_{t+1}$ against the step’s assertion $\alpha_{ij}^t$ to update the metrics (Section~\ref{sec:rq1_metrics}). After the test terminates, we store the execution trace $\tau_{ij}$ of $T$ for analysis. A test terminates when the number of steps executed by $T$ reaches $|\mathcal{A}_{ij}|$, the expected length of the corresponding test script. This termination criterion prevents unbounded execution. Tests are executed sequentially to avoid shared-state interference. For each test case, we initialize a fresh instance of the web application and restore its database to a state before the test.

\subsubsection{Results \& Discussion} Table \ref{table:rq1} summarizes the results. \tool achieves the highest TC and CT scores, both at 0.99, outperforming the best baseline by 54.7\% in TC and 28.6\% in CT.

\noindent\textbf{Why is \tool more effective?} \tool’s effectiveness stems from three reachability advantages. First, it can propose multiple actions when a test step requires them (e.g., filling multiple form fields). Second, grounding actions at the visual level using GUI grounding and SoM prompting avoids common DOM-based pitfalls such as iframes, shadow DOMs, and custom form components. Third, its two-stage action execution makes \tool more robust to out-of-distribution settings and noisy or complex UIs, enabling stronger generalization.

\noindent\textbf{Is \tool efficient?} \tool achieves a median execution time of 29 seconds and consumes a median of 10k tokens per step. It is the fastest among the compared methods, outperforming \lavague (33s), \naviqate (40s), and \pinata (38s). In terms of token consumption, it ranks second, using fewer tokens than \lavague (49k) and \pinata (19k), but slightly more than \naviqate (9k). Overall, \tool is a balance between speed and cost. This efficiency stems from its hybrid action execution design, which avoids both noisy DOM-based multi-round ranking (\naviqate) and multi-agent communication overhead (\pinata). Total computational cost scales linearly with the number of steps. Breaking down the costs by stage, token usage is dominated by Action Execution (33\%) and Page Reidentification (32\%), while execution time is primarily spent on Oracle Inference and Symbolization (34\%) and Page Reidentification (39\%). Page Reidentification is therefore the main bottleneck and a key target for future optimization.

\noindent\textbf{Is \tool maintainable?} Test actions generated by \tool may be fragile as web applications evolve. To mitigate this, \tool can cache test action and only re-invokes the action execution pipeline in Section \ref{sec:action-execution} when meaningful changes to the content/layout of the state are detected. We evaluate maintainability in a study inspired by prior work on GUI evolution~\cite{webevo}. We compare \tool with XPath-, CSS-, and Playwright-based test scripts under UI changes. The evaluation measures each method’s ability to re-identify the same GUI widgets before and after interface updates, using five tests per application: two real-world changes (on Amazon and USPS) and three synthetic changes (on BookStack) generated using transformation techniques from~\cite{guievo}. \tool preserves test actions in 39/40 cases, outperforming XPath (32/40), CSS (33/40), and Playwright (29/40).

% \begin{table}[!t]
% \caption{Results of test flow completion (RQ1) and bug detection (RQ2). -- means that the approach is not applicable for the task.} 
% \label{table:rq1_and_2}
% \centering
% \scriptsize
% \renewcommand{\arraystretch}{1.2}
% \setlength{\tabcolsep}{6pt} % slightly tighter
% \begin{tabular}{p{2.5cm} S[table-format=1.2] S[table-format=1.2] S[table-format=3.0] S[table-format=2.1] c S[table-format=1.2] S[table-format=1.2]}
% \toprule
% \multirow{2}{*}{\textbf{Approach}} & \multicolumn{4}{c}{\textbf{Test Flow Completion}} & & \multicolumn{2}{c}{\textbf{Bug Detection}} \\
% \cline{2-5}\cline{7-8}
% & \multicolumn{1}{c}{\makecell{\textbf{Task Completion} \\ \textbf{(TC)}}} 
% & \multicolumn{1}{c}{\makecell{\textbf{Correct Trace} \\ \textbf{(CT)}}} 
% & \multicolumn{1}{c}{\makecell{\textbf{Avg. Runtime} \\ \textbf{(s/step)}}} 
% & \multicolumn{1}{c}{\makecell{\textbf{Avg. Tokens} \\ \textbf{(k/step)}}} 
% &
% & \multicolumn{1}{c}{\makecell{\textbf{Precision} \\ \textbf{(\%)}}} 
% & \multicolumn{1}{c}{\makecell{\textbf{Recall} \\ \textbf{(\%)}}} \\
% \midrule
% \naviqate \cite{naviqate} & 0.10 & 0.64 & 145 & 54.6 & & \text{--} & \text{--} \\
% \lavague \cite{lavague} & 0.50 & 0.75 & 38 & 8.6 & & \text{--} & \text{--} \\
% \pinata \cite{pinata} & 0.25 & 0.71 & 48 & 19.6 & & 0.50 & 0.26 \\
% \tool & \textbf{0.84} & \textbf{0.86} & \textbf{10} & \textbf{\phantom{0}7.7} & & \textbf{0.86} & \textbf{0.80} \\
% \bottomrule
% \end{tabular}
% \end{table}

\begin{table}[!t]
\caption{Task Completion (TC) and Correct Trace (CT) across web applications
(App~\#1: BookStack, App~\#2: Indico, App~\#3: InvoiceNinja, App~\#4: PrestaShop).
\textsc{Total} denotes aggregated results across all webapps.}
\label{table:rq1}
\centering
\scriptsize
\renewcommand{\arraystretch}{1.2}
\setlength{\tabcolsep}{10pt}
\begin{tabular}{
p{1.5cm}
S[table-format=1.2] S[table-format=1.2] S[table-format=1.2] S[table-format=1.2] S[table-format=1.2]
S[table-format=1.2] S[table-format=1.2] S[table-format=1.2] S[table-format=1.2] S[table-format=1.2]
}
\toprule
\multirow{2}{*}{\textbf{Approach}} 
& \multicolumn{5}{c}{\textbf{Task Completion (TC)}} 
& \multicolumn{5}{c}{\textbf{Correct Trace (CT)}} \\
\cmidrule(r){2-6}
\cmidrule(l){7-11}
& App~\#1 & App~\#2 & App~\#3 & App~\#4 & \textsc{Total}
& App~\#1 & App~\#2 & App~\#3 & App~\#4 & \textsc{Total} \\
\midrule
\lavague
& 0.85 & 0.32 & 0.80 & 0.57 & \cellcolor{gray!10}0.64
& 0.93 & 0.49 & 0.88 & 0.78 & \cellcolor{gray!10}0.77 \\
\naviqate
& 0.78 & 0.44 & 0.60 & 0.30 & \cellcolor{gray!10}0.54
& 0.92 & 0.61 & 0.78 & 0.46 & \cellcolor{gray!10}0.70 \\
\pinata
& 0.11 & 0.04 & 0.08 & 0.09 & \cellcolor{gray!10}0.08
& 0.27 & 0.09 & 0.16 & 0.22 & \cellcolor{gray!10}0.18 \\
\tool
& \textbf{1.00} & \textbf{1.00} & \textbf{0.98} & \textbf{1.00} & \cellcolor{gray!10}\textbf{0.99}
& \textbf{1.00} & \textbf{1.00} & \textbf{0.99} & \textbf{1.00} & \cellcolor{gray!10}\textbf{0.99} \\
\bottomrule
\end{tabular}

\vspace{-1em}
\scriptsize
\end{table}

\subsection{RQ2: Bug Detection}
\label{sec:rq2}

In this section, we evaluate how well \tool and the baselines detect injected bugs in web application, using the same benchmark as in RQ1.

\subsubsection{Baselines} We exclude \lavague and \naviqate. We directly compare \tool against \pinata, the only baseline capable of bug detection.

\subsubsection{Evaluation Metrics} We use step-level outcomes. Let $s_\text{bug}$ denote the bug-injected state in a test and $\bar{s}$ the set of all other states. If $T$ generates a predicate $p(s)$, we define a true positive (TP) as $p(s_\text{bug})=\bot$, a false positive (FP) as $p(s)=\bot$ for any $s\in\bar{s}$, a false negative (FN) as $p(s_\text{bug})=\top$, and a true negative (TN) as $p(s)=\top$ for all $s\in\bar{s}$. Then $\text{precision}=\frac{|\text{TP}|}{|\text{TP}|+|\text{FP}|}$ and $\text{recall}=\frac{|\text{TP}|}{|\text{TP}|+|\text{FN}|}$. Finally, to ensure that assertions capture application behavior and to rule out coincidental matches that could be spurious TPs, we manually verify the semantic correctness of all assertions.

\subsubsection{Experiment Setup} We follow the setup in Section~\ref{sec:rq1_test_case_generation}, with two changes: (1) for each step, $T$ now generates assertion predicates $p$ that check against $\text{condition}_\text{NL}$ and $\text{expectation}_\text{NL}$. (2) $\text{bug}_{ij}$ is automatically injected into the web application at the start of each test $D_{ij}$.

\subsubsection{Results \& Discussion} Table~\ref{table:rq2} summarizes the results. \tool achieves both a precision and recall of 0.96, with absolute improvements of 0.70 and 0.27 over \pinata, respectively.

% \begin{table}[!t]
% \caption{Results of test flow completion (RQ1) and bug detection (RQ2). -- means that the approach is not applicable for the task.} 
% \label{table:rq1_and_2}
% \centering
% \scriptsize
% \renewcommand{\arraystretch}{1.2}
% \setlength{\tabcolsep}{6pt} % slightly tighter
% \begin{tabular}{p{2.5cm} S[table-format=1.2] S[table-format=1.2] S[table-format=3.0] S[table-format=2.1] c S[table-format=1.2] S[table-format=1.2]}
% \toprule
% \multirow{2}{*}{\textbf{Approach}} & \multicolumn{4}{c}{\textbf{Test Flow Completion}} & & \multicolumn{2}{c}{\textbf{Bug Detection}} \\
% \cline{2-5}\cline{7-8}
% & \multicolumn{1}{c}{\makecell{\textbf{Task Completion} \\ \textbf{(TC)}}} 
% & \multicolumn{1}{c}{\makecell{\textbf{Correct Trace} \\ \textbf{(CT)}}} 
% & \multicolumn{1}{c}{\makecell{\textbf{Avg. Runtime} \\ \textbf{(s/step)}}} 
% & \multicolumn{1}{c}{\makecell{\textbf{Avg. Tokens} \\ \textbf{(k/step)}}} 
% &
% & \multicolumn{1}{c}{\makecell{\textbf{Precision} \\ \textbf{(\%)}}} 
% & \multicolumn{1}{c}{\makecell{\textbf{Recall} \\ \textbf{(\%)}}} \\
% \midrule
% \naviqate \cite{naviqate} & 0.10 & 0.64 & 145 & 54.6 & & \text{--} & \text{--} \\
% \lavague \cite{lavague} & 0.50 & 0.75 & 38 & 8.6 & & \text{--} & \text{--} \\
% \pinata \cite{pinata} & 0.25 & 0.71 & 48 & 19.6 & & 0.50 & 0.26 \\
% \tool & \textbf{0.84} & \textbf{0.86} & \textbf{10} & \textbf{\phantom{0}7.7} & & \textbf{0.86} & \textbf{0.80} \\
% \bottomrule
% \end{tabular}
% \end{table}

\begin{table}[!t]
\caption{Precision and Recall for bug detection across web applications
(App~\#1: BookStack, App~\#2: Indico, App~\#3: InvoiceNinja, App~\#4: PrestaShop).
\textsc{Total} denotes aggregated results across all webapps.}
\label{table:rq2}
\centering
\scriptsize
\renewcommand{\arraystretch}{1.2}
\setlength{\tabcolsep}{10pt}
\begin{tabular}{
p{1.5cm}
S[table-format=1.2] S[table-format=1.2] S[table-format=1.2] S[table-format=1.2] S[table-format=1.2]
S[table-format=1.2] S[table-format=1.2] S[table-format=1.2] S[table-format=1.2] S[table-format=1.2]
}
\toprule
\multirow{2}{*}{\textbf{Approach}} 
& \multicolumn{5}{c}{\textbf{Precision}} 
& \multicolumn{5}{c}{\textbf{Recall}} \\
\cmidrule(r){2-6}
\cmidrule(l){7-11}
& App~\#1 & App~\#2 & App~\#3 & App~\#4 & \textsc{Total}
& App~\#1 & App~\#2 & App~\#3 & App~\#4 & \textsc{Total} \\
\midrule
\pinata
& 0.31 & 0.20 & 0.26 & 0.29 & \cellcolor{gray!10}0.26
& 0.70 & 0.68 & 0.68 & 0.70 & \cellcolor{gray!10}0.69 \\
\tool
& 0.98 & 0.94 & 0.94 & 1.00 & \cellcolor{gray!10}0.96
& 0.93 & 0.96 & 0.98 & 0.96 & \cellcolor{gray!10}0.96 \\
\bottomrule
\end{tabular}

\vspace{-1em}
\scriptsize
\end{table}

\noindent\textbf{Why is \tool more effective?} Analyzing execution traces $\tau$, we identify three key reasons why \tool outperforms \pinata: (1) \textit{Dynamic cross-state reasoning}: \tool can generate and track symbolic representations on the fly across multiple states. In contrast, \pinata depends on a static memory where agents must choose in advance what information to retain, and anything unrecorded is lost. This limits reasoning in tasks where crucial information is only known later or when large amounts of data make prioritization challenging. For example, in InvoiceNinja, forgetting a single detail about multiple invoices can break assertions later. (2) \textit{Exploration capability}: \tool's two-stage action execution supports robust navigation, while \pinata fails to reach certain UI elements (e.g., the timetable in Indico). (3) \textit{Full-page perception}: \tool processes the entire page screenshot at each state, whereas \pinata observes only visible elements without scrolling, potentially missing  information in long lists, tables, or grids.

\noindent\textbf{Can \tool detect real-world bugs?} We replicated 23 real-world bugs from GitHub issues. \tool detected 22 of them, compared to 15 by \pinata. The 7 bugs not detected by \pinata were due to: missing cross-state context (1), requirement misinterpretation (2), and verifier agent hallucinations on detailed pages (4). More details are available on our project page \cite{project-page}.

% \noindent\textbf{Limitations of \tool.} On this benchmark, we observe that \tool's false negatives stem from test reachability issues. False positives arise when an action fails to execute as intended.

\subsection{RQ3: Robustness Evaluation}
\label{sec:rq3}

To test whether \tool’s can generalize given all necessary information, we conduct an experiment by modifying the test requirements provided to the agent.

\subsubsection{Input Transformations} We design a set of input transformations under the assumption that all essential information (i.e., the condition, action,  expectation) are preserved. In other words, these transformations do not remove or alter the core semantics of the test case, but instead change how the information is expressed. We introduce the following four transformations:

\begin{itemize}[leftmargin=*,topsep=2pt]
    \item \textbf{Dropout}: Randomly removes 10\% of sentences to mimic incomplete requirements.
    \item \textbf{Add Noise}: Adds typos, filler or informal words to simulate casual language in communication.
    \item \textbf{Summarize}: Produces a brief, draft-style version of the test description with abbreviations.
    \item \textbf{Restyle}: Rewrites it in a different documentation style (e.g., procedural, technical, narrative).
\end{itemize}

Let the transformation functions be $f_{\text{add\_noise}}$, $f_{\text{dropout}}$, 
$f_{\text{restyle}}$, and $f_{\text{summarize}}$, each defined as 
$f : \mathcal{D} \to \mathcal{D}$, where $\mathcal{D}$ denotes the space of test requirements. In other words, given $D \in \mathcal{D}$, each transformation produces a modified test
requirement $f(D) \in \mathcal{D}$. We implement these functions by prompting LLMs to perform an initial guided transformation, followed by heuristic post-processing to produce the final output. For example, for \textbf{Add Noise}, we apply typo-generation libraries (e.g., \texttt{typo}, \texttt{nlpaug}) to introduce lexical perturbations.

\newcommand{\Transformation}[3]{%
\begin{minipage}[t]{0.24\textwidth}
\begin{tcolorbox}[
    title=#1,
    colback=#2!5!white,
    colframe=#2!75!black,
    height=2.2cm,
    valign=center,
    boxsep=1mm,
    left=1mm,
    right=1mm,
    top=1mm,
    bottom=1mm,
    fontupper=\footnotesize,
    fonttitle=\footnotesize
]
#3
\end{tcolorbox}
\end{minipage}%
}

\begin{figure}[h]
\noindent
\Transformation{Add Noise}{red}{frmo tr dsboard amd tpapin9 the "Boosk" ljnk nestled withi he naviatno mneu [...]}
\hfill
\Transformation{Dropout}{orange}{\textit{Redacted}. Books listing page appears. Verify "Create New Book" link is visible [...]}
\hfill
\Transformation{Restyle}{violet}{To begin your journey through the digital library, start by navigating [...]}
\hfill
\Transformation{Summarize}{olive}{From dash, click 'Books' -> verify 'Create New Book' link -> click it -> form opens [...]}
\caption{Example of transformed test requirements. Original text: ``\textit{Click "Books" link in navigation. Books listing page appears. Verify "Create New Book" link is visible [...]}''}
\end{figure}

\subsubsection{Model Selection} 
Beyond \tool’s base model (GPT-4.1), we evaluate four open-source Qwen2.5-VL models (72B, 32B, 7B, and 3B). Qwen2.5-VL is trained with GUI grounding data, and has shown strong generalization on GUI testing and agentic benchmarks~\cite{guitestingarena,qwen,wang2025mmbench}.

\subsubsection{Experiment Setup} We follow the same setup as in Section~\ref{sec:rq2}, with the difference that, for each test, the input test requirement $D_{ij}$ is first transformed using the transformation functions: $f_{\text{default}}$, $f_{\text{add\_noise}}$, $f_{\text{dropout}}$, $f_{\text{restyle}}$, and $f_{\text{summarize}}$. We use metrics defined in Section~\ref{sec:rq1_test_case_generation} and~\ref{sec:rq2}.

\subsubsection{Results \& Discussion}

% Table \ref{table:rq3_rq4} shows the results. No single transformation dominates across models. Averaging the performance drop by transformation yields a mean of 44.5\% ($\sigma$ = 7.2\%), with the largest drop caused by “Add Noise” at 55.2\%. Stylistic transformations are largely stable, which suggests that LLMs can reliably handle input parsing when tasked with inferring and extracting information, rather than filtering out noise. 

% When comparing task completion and correct trace across models and transformations, we find that the gap between them is both small and stable (mean discrepancy = 0.15, variance = 0.09). It shows that failures are not caused by randomness or inconsistent execution of actions (poor stability), rather it points to the model's reliability in parsing inputs. Small parsing errors, such as omitting or misordering a step, can break a test. Although GPT-4.1 remains the strongest overall, larger parameter sizes do not guarantee better parsing. For instance, Qwen-2.5VL-7B outperforms Qwen-2.5VL-32B. Closer inspection shows that the latter parses instructions verbatim but misses some steps, while the former compresses them into simplified [action] [noun] forms that capture all steps. It is the parsing strategy, not size alone, that is the main factor of performance.
Table \ref{table:rq3_and_4} shows the results. We observe that no transformation consistently reduces TC or CT for every model, indicating the absence of a universal “worst-case” transformation. For instance, DO reduces TC for Qwen2.5-VL-32b from 0.65 (DF) to 0.63, while Qwen2.5-VL-7b remains largely unaffected at 0.70. This suggests that each model exhibits its own strengths and weaknesses, reacting differently to various transformations: GPT-4.1 maintains high performance under noise (AN = 0.93) but is more impacted by DO, RS, and SU (0.70 each), whereas smaller models like Qwen2.5-VL-3b are highly sensitive to noise (AN = 0.48, SU = 0.41).

In general, performance declines as model size decreases, but the decline is not uniform. Initially, reducing model size by roughly half results in modest performance drops of less than 10\%. However, 7B is a critical threshold where TC and CT start to decline sharply by 20–30\%. Thus, we suggest that for cost considerations, 7B models may serve as the minimum viable option, whereas for performance, local models should be at least 72B parameters to reliably match or exceed GPT-4.1.

Finally, there is a noticeable gap between TC and CT, with differences ranging from 0.05 to 0.09 across models. We observe that models can omit or introduce redundant steps in transformed test requirements, leading to errors in trace execution. For example, details about filling a form (how many fields, what expected input) are not interpreted correctly during parsing.

% No single transformation dominates across models. Averaging the performance drop by transformation yields 44.5\% ($\sigma$ = 7.2\%), with the largest drop from “Add Noise” at 55.2\%. Stylistic transformations remain largely stable, which suggests that LLMs can reliably handle input parsing when tasked with inferring and extracting information, rather than filtering out noise. Comparing task completion and correct trace, the gap is small and stable (mean discrepancy = 0.15, variance = 0.09), indicating failures stem from parsing errors rather than execution instability. Small parsing errors, like omitted or misordered steps, can break tests. GPT-4.1 performs best overall, but larger model size does not guarantee better parsing; for example, Qwen-2.5VL-7B outperforms Qwen-2.5VL-32B by compressing instructions into simplified [action][noun] sequences, capturing all steps.

There are two key takeaways. First, all necessary information must be present in the requirements, as accurate input parsing is the strongest predictor of downstream task performance. Second, style and formatting variations can be overcome by designing specialized semantic parsers tailored to the specific domain and language style (e.g., PRD parsers, email parsers, or Slack/chat message parsers) that restructure inputs into step sequences that are correct, complete, and concise.

% \begin{table}[!t]
% \caption{Results of robustness evaluation (RQ3) and model comparison (RQ4). \textbf{Bold} indicates the best-performing model; \underline{underline} indicates the second-best. \textbf{TC:} Task Completion, \textbf{CT:} Correct Trace.}
% \label{table:rq3_rq4}
% \centering
% \scriptsize
% \renewcommand{\arraystretch}{1.2}
% \setlength{\tabcolsep}{9pt}
% \begin{tabular}{p{2.5cm} cc cc cc cc cc cc}
% \toprule
% \textbf{Model} 
% & \multicolumn{2}{c}{\textbf{Default}} 
% & \multicolumn{2}{c}{\textbf{Add Noise}}
% & \multicolumn{2}{c}{\textbf{Dropout}} 
% & \multicolumn{2}{c}{\textbf{Restyle}} 
% & \multicolumn{2}{c}{\textbf{Summarize}}\\
% \cmidrule(lr){2-5} \cmidrule(lr){6-7} \cmidrule(lr){8-9} \cmidrule(lr){10-11} \cmidrule(lr){12-13}
% & TC & CT & TC & CT & TC & CT & TC & CT & TC & CT \\
% \midrule
% GPT-4.1 & \underline{0.67} & \textbf{0.85} & \underline{0.60} & \underline{0.75} & \underline{0.80} & 0.83 & \textbf{0.67} & \textbf{0.85} & \underline{0.60} & \underline{0.75}\\
% Qwen2.5-VL-72b & \underline{0.67} & \underline{0.69} & \textbf{0.67} & \textbf{0.92} & 0.67 & \underline{0.84} & 0.50 & 0.69 & \textbf{0.67} & \textbf{0.92}\\
% Qwen2.5-VL-36b & 0.33 & 0.46 & 0.33 & 0.69 & 0.33 & 0.69 & \underline{0.67} & \underline{0.84} & \textbf{0.67} & 0.54\\
% Qwen2.5-VL-7b & \textbf{0.83} & \textbf{0.85} & 0.50 & 0.61 & \textbf{0.83} & \textbf{0.85} & 0.50 & \underline{0.84} & 0.50 & 0.54\\
% Qwen2.5-VL-3b & 0.33 & 0.46 & 0.17 & 0.31 & 0.17 & 0.31 & 0.33 & 0.61 & 0.50 & 0.54\\
% \bottomrule
% \end{tabular}
% \end{table}

\begin{table}[!t]
\caption{Task Completion (TC) and Correct Trace (CT) across test requirement transformations, evaluated using \tool.
\textsc{Total} denotes aggregated results across all transformations.
Abbreviations: DF = Default, AN = Add Noise, DO = Dropout, RS = Restyle, SU = Summarize.}
\label{table:rq3_and_4}
\centering
\scriptsize
\renewcommand{\arraystretch}{1.2}
\setlength{\tabcolsep}{7pt} % tighter spacing
\begin{tabular}{p{2cm}
S[table-format=1.2] S[table-format=1.2] S[table-format=1.2] S[table-format=1.2] S[table-format=1.2] S[table-format=1.2]
S[table-format=1.2] S[table-format=1.2] S[table-format=1.2] S[table-format=1.2] S[table-format=1.2] S[table-format=1.2]
}
\toprule
\multirow{2}{*}{\shortstack[l]{\textbf{Model}}}
& \multicolumn{6}{c}{\textbf{Task Completion (TC)}}
& \multicolumn{6}{c}{\textbf{Correct Trace (CT)}} \\
\cmidrule(r){2-7} \cmidrule(l){8-13}
& \textbf{DF} & \textbf{AN} & \textbf{DO} & \textbf{RS} & \textbf{SU} & \textsc{Total}
& \textbf{DF} & \textbf{AN} & \textbf{DO} & \textbf{RS} & \textbf{SU} & \textsc{Total} \\
\midrule
\textsc{GPT-4.1}       & 1.00 & 0.93 & 0.70 & 0.70 & 0.70 & \cellcolor{gray!10}0.81
          & 1.00 & 0.95 & 0.81 & 0.78 & 0.78 & \cellcolor{gray!10}0.86 \\
\textsc{Qwen2.5-VL-72b}  & 1.00 & 0.93 & 0.85 & 0.70 & 0.81 & \cellcolor{gray!10}0.85
          & 1.00 & 0.95 & 0.89 & 0.82 & 0.85 & \cellcolor{gray!10}0.90 \\
\textsc{Qwen2.5-VL-32b}  & 0.65 & 0.89 & 0.63 & 0.78 & 0.81 & \cellcolor{gray!10}0.75
          & 0.76 & 0.94 & 0.76 & 0.85 & 0.88 & \cellcolor{gray!10}0.84 \\
\textsc{Qwen2.5-VL-7b}   & 1.00 & 0.74 & 0.70 & 0.70 & 0.63 & \cellcolor{gray!10}0.72
          & 1.00 & 0.84 & 0.84 & 0.83 & 0.74 & \cellcolor{gray!10}0.83 \\
\textsc{Qwen2.5-VL-3b}   & 1.00 & 0.48 & 0.52 & 0.70 & 0.41 & \cellcolor{gray!10}0.57
          & 1.00 & 0.61 & 0.66 & 0.77 & 0.47 & \cellcolor{gray!10}0.66 \\
\bottomrule
\end{tabular}

\vspace{-1em}
\scriptsize
\end{table}

\begin{figure}[t]
    \centering
    % First figure
    \begin{subfigure}[b]{0.49\textwidth}
        \centering
        \includegraphics[trim=0 10 0 5, clip, width=\textwidth]{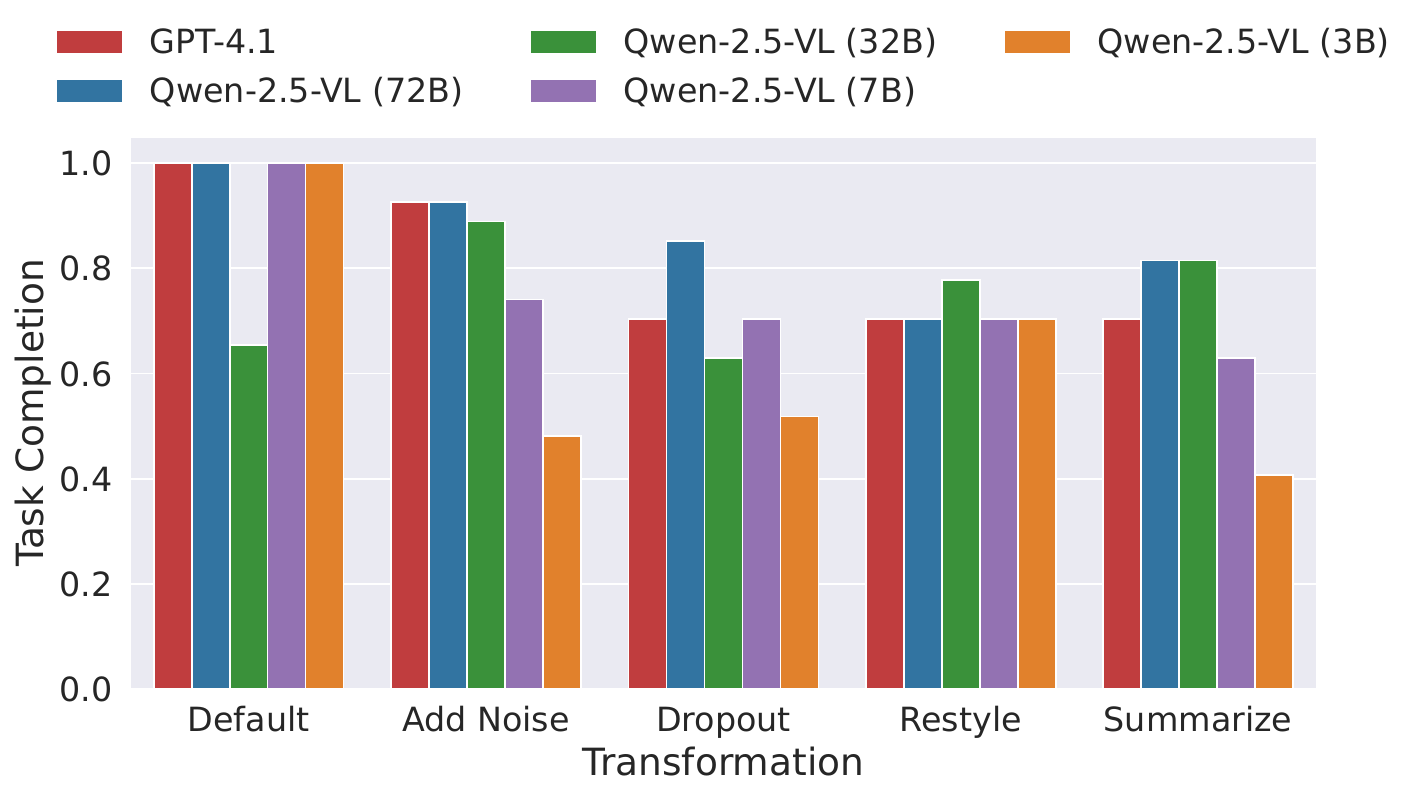}
        \caption{Task Completion}
        \label{fig:rq3_task_completion}
    \end{subfigure}
    \hfill
    % Second figure
    \begin{subfigure}[b]{0.49\textwidth}
        \centering
        \includegraphics[trim=0 10 0 5, clip, width=\textwidth]{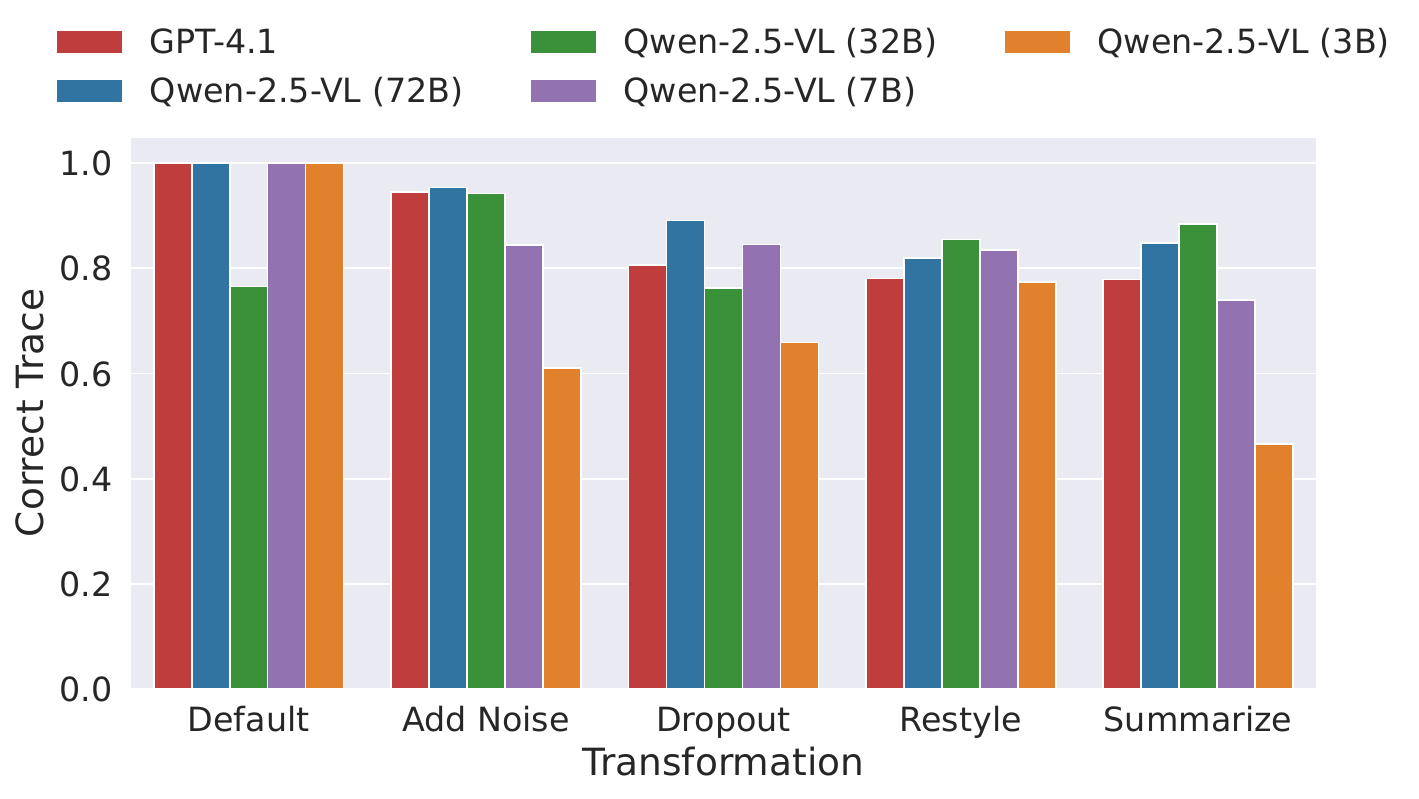}
        \caption{Correct Trace}
        \label{fig:rq3_4_correct_trace}
    \end{subfigure}
    
    \caption{Performance of different models (RQ4) under different transformed input requirements (RQ3).}
    \label{fig:side_by_side}
    \vspace{-5mm}
\end{figure}
\subsection{RQ4: Model Comparison}

We perform an ablation study of \tool by replacing its base model with alternative LLMs of varying sizes and capacities.

\subsubsection{Experiment Setup} We follow the setup and metrics in Section~\ref{sec:rq3}, but we evaluate only \tool and vary its underlying model on the benchmark without transformation.

\subsubsection{Results \& Discussion} Table \ref{table:rq3_and_4} (default, DF column) shows that model performance is generally stable. However, the picture changes when models generate predicate assertions.

\noindent\textbf{Assertion Quality.} GPT-4.1 shows strong DSL usage: 91\% of predicate assertions reference both prior and current states, 8\% only the current state, and 0.2\% non-adjacent states. Of declared symbols, 41.3\% represent physical concepts (e.g., \texttt{Cart}) and 58.7\%  UI components (e.g., \texttt{DropDown}). Most assertions perform existence checks (membership 78\%, \texttt{is None} 14\%, list length 18\%) and 51\% use relational comparisons. Common built-ins include \texttt{len}, \texttt{any}, \texttt{set}, \texttt{all}, \texttt{next}, and \texttt{reversed}.

\noindent\textbf{Analysis of Assertion Errors.}
Challenges occur in local models. On proprietary models (e.g., GPT-4.1), failures can be mitigated through prompt tuning. For local models (Qwen-2.5VL series), however, we observed the following issues despite prompt tuning:
\begin{itemize}[leftmargin=*,topsep=2pt]
    \item \textit{Incorrect symbol declaration or usage (52\% of cases):} Symbols declared but unused, or used without declaration. Some models treat \texttt{BaseModel} as a concrete symbol rather than an abstract schema (akin to using an abstract class), or misuse symbols to extract non-visual data (e.g., HTML).
    \item \textit{Incorrect usage of the assertion DSL in Oracle Inference (39\% of cases):} Hallucinated attributes, imports (e.g., \texttt{state.isNotificationEnabled}) and method calls (e.g., \texttt{session.extract()}).
    \item \textit{Runtime errors in Oracle Execution (9\% of cases):} Mismatched data types in equality comparisons and incorrect membership checks (e.g., \texttt{in}/\texttt{not in} applied to non-set data).
\end{itemize}

To improve local model performance, we recommend: (1) fine-tuning with DSL examples, (2) providing access to DSL references via an external source (e.g., retrieval-augmented generation), or (3) constraining outputs to be syntactically correct according to the DSL. 

% shows that Qwen2.5-VL-7b excels at task completion, thanks to its step pre-processing into simplified [action] [noun] descriptions, which improves navigation. Larger models, however, are more reliable at producing runnable DSL-compliant assertions. Assertion failure rates rise steeply from 0.04 (GPT-4.1) to 0.15, 0.27, 0.29, and 0.66 for Qwen2.5-VL-72b, -32b, -7b, and -3b, respectively. For consistent requirement checking, models $\geq72$b offer the best trade-off between cost and effectiveness. Runtime errors stem from three main issues: (1) smaller models ($\leq7$b) often use undefined symbols, (2) mid-sized models ($7$–$32$b) hallucinate methods or attributes, and (3) larger models ($\geq32$b) misuse assertions (e.g., applying membership check on state). In short, navigation favors smaller models, while assertions favor larger ones.
\section{Case Study}
\noindent\textbf{Setup.} In addition to our empirical studies, 
we collaborated with China Mobile on its no-code platform $P$,
which supports relational data modeling and drag-and-drop UI design for enterprise applications.
Its target users are non-technical staff who build internal enterprise applications. 
We were granted access to an in-progress warehouse management system $w$, and converted its PRD into individual requirements $d$ for \tool to check consistency against $w$.
In total, \tool uncovered eight bugs (Table \ref{tab:case-study-bugs}).

\noindent\textbf{Results.} Of the eight bugs, five (62.5\%) were data-binding issues, two were UI issues, and one was a navigation issue, demonstrating \tool’s strength in detecting data-related bugs often missed by baselines. Technically, 
%\lavague and \naviqate can only detect the navigation bug (1/8), while 
\pinata is limited to detecting UI and navigation issues (3/8) and would catch data issues only if explicitly specified. The PRD was pre-processed into 56 test inputs (6.4 mins), and all tests ran in 32.7 mins. \tool’s mean time to detect (MTTD) was 4.9 mins, with a defect density of 0.14 bugs per page. These results show that \tool provides both practical effectiveness and testing efficiency in real-world applications.

% \noindent\textbf{Results.} Of the eight bugs, five (62.5\%) are data-binding issues, two are UI issues, and one is a navigation issue, highlighting \tool’s strength in detecting data-related bugs often missed by baselines. For comparison, \lavague and \naviqate detect only navigation bugs (1/8), while \pinata detects UI and navigation issues (3/8) and would catch data issues only if explicitly specified. We pre-processed the PRD into 56 test inputs (6.4 mins) and ran all tests in 32.7 mins. \tool detected eight bugs, yielding a mean time to detect (MTTD) of 4.9 mins and a defect density of 0.14 bugs per page. These results demonstrate that \tool delivers both practical effectiveness and testing efficiency in real-world applications.
% test time = (3 * 2 * 14 + 4 * 2 * 14) * time per step
% update create is 4 steps, read delete 3
% 51 pages
\begin{table}[!t]
\centering
\caption{Bugs detected by \tool in the case study.}
\label{tab:case-study-bugs}
\scriptsize
\renewcommand{\arraystretch}{1.2}
\setlength{\tabcolsep}{4pt}
\begin{tabular}{c|l|l|l|l|p{5cm}}
\hline
\textbf{\#} & \textbf{Section} & \textbf{Page} & \textbf{Feature/Action} & \textbf{Bug Type} & \textbf{Bug Description} \\ \hline
1 & \multirow[t]{3}{*}{Warehouses} & Warehouse Info & Table & Data & Some required fields in the table are empty. \\
2 & & Storage Area & Create & UI & Duplicate ``Warehouse Name'' form fields. \\
3 & & Storage Unit & Search & Data & Dropdown options for "Warehouse" is inconsistent with available warehouses. \\ \hline
4 & Receipts & Receipt Info & Table & Nav & Clicking ``Details'' leads to an error page. \\ \hline
5 & \multirow[t]{4}{*}{Assets} & Inventory & Search Form & Data & Dropdown options for "Warehouse" are not bound to the actual table data. \\
6 &  & Inventory & Search Form & Data & Dropdown options for "Storage Area" are not bound to the actual table data. \\
7 &  & Asset & Search Form & Data & Dropdown options for "Supplier" is empty.\\ \hline
8 & Device Management & Cameras & Search Form & UI & Query field names are incorrect. \\ \hline
\end{tabular}
\end{table}

\section{Discussion}

\noindent\textbf{Threats to Validity.} Internally, our metrics may underestimate tester performance, as multiple paths can achieve the same functionality. Future work could consider final page layout, content, or application state as additional indicators. Externally, our benchmark (webapp, test case, bugs) may not fully reflect real-world scenarios. However, since \tool models testing as a consistency problem using a Pythonic DSL, it can handle any bug that causes behavior to diverge from requirements. A final limitation is that we assume requirements are self-contained and complete, specifying all conditions, actions, and expected outcomes in order.

\noindent\textbf{Semantic Parsing for Specification-based Testing.} Following our discussion above, experiments show that input parsing strategy and quality drive performance. Natural language requirements are often ambiguous, incomplete, or context-dependent. Parsing requires semantic understanding and pre-processing, not just extraction. LLM agents must act as proactive semantic parsers, transforming requirements into machine-understandable, executable representations. This includes identifying ambiguities, ask the user clarifying questions, and retrieving context from an external knowledge base where needed.
\section{Related Work}

\subsection{Automated GUI Testing}

Automated GUI testing simulates user interactions (e.g., clicks) to validate application functionality via its GUI. Random techniques explore the AUT by fuzzing random actions (e.g., Monkey \cite{Monkey}, Gremlins.js \cite{gremlin.js}) or by randomly interacting with detected widgets (White et al. \cite{white2019improving}). Model-based approaches (e.g., Crawljax \cite{crawljax}, ATUSA \cite{atusa}, Stoat \cite{stoat}) construct navigational or behavioral models (e.g., flow graphs, state machines) of the AUT and derive test cases from them. To prune redundant model states, works like Judge \cite{judge}, WebEmbed \cite{webembed}, Corazza et al. \cite{corazza2021web}, FragGen \cite{fraggen}, and NDStudy \cite{ndstudy} detect and remove near-duplicate states. Systemic strategies try to generate test cases that optimizes a test objective (e.g., code coverage), which can be done through search-based techniques (e.g., DIG \cite{dig}, SubWeb \cite{subweb}, FeedEx \cite{feedex}, RoboTest \cite{robotest}, Sapienz \cite{sapienz}, TimeMachine \cite{timemachine}) and symbolic execution (e.g., Apollo \cite{apollo}). Reinforcement learning (RL) approaches frame testing as a sequential decision problem using Q-learning or policy optimization to guide exploration on the AUT (e.g., AutoBlackTest \cite{autoblacktest}, QExplore \cite{qexplore}, WebExplor \cite{webexplor}, WebQT \cite{webqt}, WebRLED \cite{webrled}, UniRLTest \cite{unirl}, PIRL-Test \cite{pirltest}, Hawkeye \cite{hawkeye}). These exploration-based methods prioritize coverage over requirements. Specification-based testing uses requirements for targeted validation of user flows. Kea \cite{kea} uses a property description language to manually specify properties for Android apps. In contrast, \tool automatically derives symbolic assertions from rich contextual test information. Complementary works improve test efficiency \cite{stile} and stability \cite{liu2024wefix,pei2025non,zhang2024towards}.

\subsection{LLM for GUI Testing}

\noindent\textbf{Input Generation.} \textsc{QTypist} \cite{QTypist} produces context-aware text inputs for realistic testing. \textsc{InputBlaster} \cite{InputBlaster} mutates input strings to trigger crashes, and \textsc{FormNexus} \cite{FormNexus} validates form functionality via constraint-based testing. These approaches improve E2E testing reachability.

\noindent\textbf{Mobile Applications.} Several works, such as \textsc{GPTDroid}\cite{GPTDroid,GPTDroid-v2}, \textsc{DroidAgent}\cite{DroidAgent}, \textsc{LLMDroid}, \textsc{Guardian}\cite{Guardian}, \textsc{AUITestAgent}\cite{AUITestAgent}, \textsc{Trident}\cite{Trident}, \textsc{A11yScan}\cite{a11yscan} and \textsc{XUAT-Copilot}\cite{xuat-copilot}, focus on mobile E2E testing, using techniques like functionality-aware dialogues, coverage-guided exploration, multi-agent planning, and verification inference. Garcia et al.~\cite{garcia2024use} also study how testers collaborate with LLMs in mobile testing. These works target mobile instead of web platforms.

% \noindent\textbf{Mobile Applications.} \textsc{GPTDroid} \cite{GPTDroid,GPTDroid-v2} converts GUI information into textual prompts and conducts functionality-aware dialogues with an LLM to determine the next action. \textsc{DroidAgent} \cite{DroidAgent} generates high-level task goals and attempts to achieve them through LLM-guided interaction. \textsc{LLMDroid} incorporates code coverage feedback to guide exploration, while \textsc{Guardian} \cite{Guardian} improves robustness by refining the UI action space and replanning when assumptions are invalidated. \textsc{AUITestAgent} \cite{AUITestAgent} uses LLMs to infer verification points from interaction logs and UI structures. \textsc{Trident} \cite{Trident} and \textsc{XUAT-Copilot} \cite{xuat-copilot} introduce multi-agent frameworks to handle planning, execution, and verification jointly. Garcia et al. \cite{garcia2024use} conduct a user study on how testers collaborate with LLMs during mobile E2E testing. These systems operate on mobile platforms and are not directly comparable to our work.

\noindent\textbf{Web Applications.} Zimmermann et al. \cite{zimmermann2023gui} and VETL \cite{wang2024leveraging} propose the first LLM and multimodal LLM-based GUI testing agent, respectively. \textsc{AutoAUT} \cite{autoblacktest} and Leotta et al. \cite{leotta2024ai} conduct feasibility studies and user interviews to understand how LLMs can support acceptance testing workflows. \textsc{AxNav} \cite{Axnav} and \textsc{UXAgent} \cite{UXAgent} target accessibility and usability testing, respectively. These tools do not perform full E2E flow validation. \textsc{AutoE2E} \cite{autoe2e} and \textsc{Temac} \cite{Temac} infer features from the application under test (AUT) and use them to drive test case generation. \textsc{LLM-Explorer} \cite{LLM-Explorer} maintains an abstract UI state and interaction graph to guide exploration. However, these systems primarily target coverage and do not verify expected outcomes. \naviqate \cite{naviqate} ranks actionable elements by relevance to a goal to guide interaction, but does not verify whether the final outcome satisfies the user objective. In summary, existing LLM-based web testers are limited oracles that focus on end states or explicit requirements, missing inconsistencies not captured in the specification. \tool addresses this with formalized test specifications and pre/post-condition verification, enabling stable and reliable testing that also accounts for inferred implicit requirements.

\section{Conclusion}

In this work, we show that LLM agents, when paired with symbolic modeling and a DSL for formalized assertions, can serve as reliable automated GUI testers. We propose \tool,
%which bridges the gap between LLM agents and specification-based testing by 
which detects implicit, context-dependent bugs with high precision and recall, while remaining robust across diverse inputs and model scales.
\section*{Data Availability}

Our benchmark, the source code of \tool and baselines, and all scripts for setting up and running experiments are available at \url{https://github.com/code-philia/WebTestPilot}. For more details (i.e., prompts, case study, etc.), please visit \url{https://sites.google.com/view/webtestpilot}.
\section*{Acknowledgement}

We thank the reviewers for their constructive feedback and Haozhe Wei for his contributions to the benchmark construction. This research is conducted in collaboration with China Mobile, and is supported in part by the National Natural Science Fundation of China (62572300), the Minister of Education, Singapore (MOE-T2EP20124-0017, MOET32020-0004), the National Research Foundation, Singapore and the Cyber Security Agency under its National Cybersecurity R\&D Programme (NCRP25-P04-TAICeN), DSO National Laboratories under the AI Singapore Programme (AISG Award No: AISG2-GC-2023-008-1B), and Cyber Security Agency of Singapore under its National Cybersecurity R\&D Programme and CyberSG R\&D Cyber Research Programme Office. Any opinions, findings and conclusions or recommendations expressed in this material are those of the author(s) and do not reflect the views of National Research Foundation, Singapore, Cyber Security Agency of Singapore as well as CyberSG R\&D Programme Office, Singapore.

\bibliographystyle{ACM-Reference-Format}
\bibliography{bibliography}


\begin{thebibliography}{82}

%%% the Web address.

\ifx \showCODEN    \undefined \def \showCODEN     #1{\unskip}     \fi
\ifx \showDOI      \undefined \def \showDOI       #1{#1}\fi
\ifx \showISBNx    \undefined \def \showISBNx     #1{\unskip}     \fi
\ifx \showISBNxiii \undefined \def \showISBNxiii  #1{\unskip}     \fi
\ifx \showISSN     \undefined \def \showISSN      #1{\unskip}     \fi
\ifx \showLCCN     \undefined \def \showLCCN      #1{\unskip}     \fi
\ifx \shownote     \undefined \def \shownote      #1{#1}          \fi
\ifx \showarticletitle \undefined \def \showarticletitle #1{#1}   \fi
\ifx \showURL      \undefined \def \showURL       {\relax}        \fi
% invisible to TeX
\providecommand\bibfield[2]{#2}
\providecommand\bibinfo[2]{#2}
\providecommand\natexlab[1]{#1}
\providecommand\showeprint[2][]{arXiv:#2}

\bibitem[Alian et~al\mbox{.}(2024)]%
        {FormNexus}
\bibfield{author}{\bibinfo{person}{Parsa Alian}, \bibinfo{person}{Noor Nashid}, \bibinfo{person}{Mobina Shahbandeh}, {and} \bibinfo{person}{Ali Mesbah}.} \bibinfo{year}{2024}\natexlab{}.
\newblock \showarticletitle{Semantic constraint inference for web form test generation}. In \bibinfo{booktitle}{\emph{Proceedings of the 33rd ACM SIGSOFT International Symposium on Software Testing and Analysis}}. \bibinfo{pages}{932--944}.
\newblock


\bibitem[Alian et~al\mbox{.}(2025)]%
        {autoe2e}
\bibfield{author}{\bibinfo{person}{Parsa Alian}, \bibinfo{person}{Noor Nashid}, \bibinfo{person}{Mobina Shahbandeh}, \bibinfo{person}{Taha Shabani}, {and} \bibinfo{person}{Ali Mesbah}.} \bibinfo{year}{2025}\natexlab{}.
\newblock \showarticletitle{{ Feature-Driven End-to-End Test Generation }}.
\newblock \bibinfo{journal}{\emph{2025 IEEE/ACM 47th International Conference on Software Engineering (ICSE)}} (\bibinfo{year}{2025}), \bibinfo{pages}{450--462}.
\newblock
\urldef\tempurl%
\url{https://doi.org/10.1109/ICSE55347.2025.00141}
\showDOI{\tempurl}


\bibitem[Artzi et~al\mbox{.}(2008)]%
        {apollo}
\bibfield{author}{\bibinfo{person}{Shay Artzi}, \bibinfo{person}{Adam Kiezun}, \bibinfo{person}{Julian Dolby}, \bibinfo{person}{Frank Tip}, \bibinfo{person}{Danny Dig}, \bibinfo{person}{Amit Paradkar}, {and} \bibinfo{person}{Michael~D Ernst}.} \bibinfo{year}{2008}\natexlab{}.
\newblock \showarticletitle{Finding bugs in dynamic web applications}. In \bibinfo{booktitle}{\emph{Proceedings of the 2008 international symposium on Software testing and analysis}}. \bibinfo{pages}{261--272}.
\newblock


\bibitem[Bai et~al\mbox{.}(2025)]%
        {qwen}
\bibfield{author}{\bibinfo{person}{Shuai Bai}, \bibinfo{person}{Keqin Chen}, \bibinfo{person}{Xuejing Liu}, \bibinfo{person}{Jialin Wang}, \bibinfo{person}{Wenbin Ge}, \bibinfo{person}{Sibo Song}, \bibinfo{person}{Kai Dang}, \bibinfo{person}{Peng Wang}, \bibinfo{person}{Shijie Wang}, \bibinfo{person}{Jun Tang}, {et~al\mbox{.}}} \bibinfo{year}{2025}\natexlab{}.
\newblock \showarticletitle{Qwen2. 5-vl technical report}.
\newblock \bibinfo{journal}{\emph{arXiv preprint arXiv:2502.13923}} (\bibinfo{year}{2025}).
\newblock


\bibitem[Baral et~al\mbox{.}(2024)]%
        {baral2024automating}
\bibfield{author}{\bibinfo{person}{Kesina Baral}, \bibinfo{person}{John Johnson}, \bibinfo{person}{Junayed Mahmud}, \bibinfo{person}{Sabiha Salma}, \bibinfo{person}{Mattia Fazzini}, \bibinfo{person}{Julia Rubin}, \bibinfo{person}{Jeff Offutt}, {and} \bibinfo{person}{Kevin Moran}.} \bibinfo{year}{2024}\natexlab{}.
\newblock \showarticletitle{Automating gui-based test oracles for mobile apps}. In \bibinfo{booktitle}{\emph{Proceedings of the 21st International Conference on Mining Software Repositories}}. \bibinfo{pages}{309--321}.
\newblock


\bibitem[Biagiola et~al\mbox{.}(2017)]%
        {subweb}
\bibfield{author}{\bibinfo{person}{Matteo Biagiola}, \bibinfo{person}{Filippo Ricca}, {and} \bibinfo{person}{Paolo Tonella}.} \bibinfo{year}{2017}\natexlab{}.
\newblock \showarticletitle{Search based path and input data generation for web application testing}. In \bibinfo{booktitle}{\emph{International Symposium on Search Based Software Engineering}}. Springer, \bibinfo{pages}{18--32}.
\newblock


\bibitem[Biagiola et~al\mbox{.}(2019)]%
        {dig}
\bibfield{author}{\bibinfo{person}{Matteo Biagiola}, \bibinfo{person}{Andrea Stocco}, \bibinfo{person}{Filippo Ricca}, {and} \bibinfo{person}{Paolo Tonella}.} \bibinfo{year}{2019}\natexlab{}.
\newblock \showarticletitle{Diversity-based web test generation}. In \bibinfo{booktitle}{\emph{Proceedings of the 2019 27th ACM Joint Meeting on European Software Engineering Conference and Symposium on the Foundations of Software Engineering}}. \bibinfo{pages}{142--153}.
\newblock


\bibitem[Bookstack(2015)]%
        {bookstack}
Bookstack \bibinfo{year}{2015}\natexlab{}.
\newblock \bibinfo{howpublished}{\url{https://github.com/BookStackApp/BookStack}}.
\newblock


\bibitem[Chang et~al\mbox{.}(2023)]%
        {webqt}
\bibfield{author}{\bibinfo{person}{Xiaoning Chang}, \bibinfo{person}{Zheheng Liang}, \bibinfo{person}{Yifei Zhang}, \bibinfo{person}{Lei Cui}, \bibinfo{person}{Zhenyue Long}, \bibinfo{person}{Guoquan Wu}, \bibinfo{person}{Yu Gao}, \bibinfo{person}{Wei Chen}, \bibinfo{person}{Jun Wei}, {and} \bibinfo{person}{Tao Huang}.} \bibinfo{year}{2023}\natexlab{}.
\newblock \showarticletitle{A reinforcement learning approach to generating test cases for web applications}. In \bibinfo{booktitle}{\emph{2023 IEEE/ACM International Conference on Automation of Software Test (AST)}}. IEEE, \bibinfo{pages}{13--23}.
\newblock


\bibitem[Chevrot et~al\mbox{.}(2025)]%
        {pinata}
\bibfield{author}{\bibinfo{person}{Antoine Chevrot}, \bibinfo{person}{Alexandre Vernotte}, \bibinfo{person}{Jean-R{\'e}my Falleri}, \bibinfo{person}{Xavier Blanc}, \bibinfo{person}{Bruno Legeard}, {and} \bibinfo{person}{Aymeric Cretin}.} \bibinfo{year}{2025}\natexlab{}.
\newblock \showarticletitle{Are Autonomous Web Agents Good Testers?}
\newblock \bibinfo{journal}{\emph{Proceedings of the ACM on Software Engineering}} \bibinfo{volume}{2}, \bibinfo{number}{ISSTA} (\bibinfo{year}{2025}), \bibinfo{pages}{206--228}.
\newblock


\bibitem[Corazza et~al\mbox{.}(2021)]%
        {corazza2021web}
\bibfield{author}{\bibinfo{person}{Anna Corazza}, \bibinfo{person}{Sergio Di~Martino}, \bibinfo{person}{Adriano Peron}, {and} \bibinfo{person}{Luigi Libero~Lucio Starace}.} \bibinfo{year}{2021}\natexlab{}.
\newblock \showarticletitle{Web application testing: Using tree kernels to detect near-duplicate states in automated model inference}. In \bibinfo{booktitle}{\emph{Proceedings of the 15th ACM/IEEE International Symposium on Empirical Software Engineering and Measurement (ESEM)}}. \bibinfo{pages}{1--6}.
\newblock


\bibitem[Cucumber(2014)]%
        {cucumber}
Cucumber \bibinfo{year}{2014}\natexlab{}.
\newblock
\newblock
\urldef\tempurl%
\url{https://cucumber.io/}
\showURL{%
\tempurl}


\bibitem[Di~Meglio et~al\mbox{.}(2025)]%
        {e2egit}
\bibfield{author}{\bibinfo{person}{Sergio Di~Meglio}, \bibinfo{person}{Luigi Libero~Lucio Starace}, \bibinfo{person}{Valeria Pontillo}, \bibinfo{person}{Ruben Opdebeeck}, \bibinfo{person}{Coen De~Roover}, {and} \bibinfo{person}{Sergio Di~Martino}.} \bibinfo{year}{2025}\natexlab{}.
\newblock \showarticletitle{E2EGit: A Dataset of End-to-End Web Tests in Open Source Projects}. In \bibinfo{booktitle}{\emph{2025 IEEE/ACM 22nd International Conference on Mining Software Repositories (MSR)}}. IEEE, \bibinfo{pages}{836--840}.
\newblock


\bibitem[Dong et~al\mbox{.}(2020)]%
        {timemachine}
\bibfield{author}{\bibinfo{person}{Zhen Dong}, \bibinfo{person}{Marcel B{\"o}hme}, \bibinfo{person}{Lucia Cojocaru}, {and} \bibinfo{person}{Abhik Roychoudhury}.} \bibinfo{year}{2020}\natexlab{}.
\newblock \showarticletitle{Time-travel testing of android apps}. In \bibinfo{booktitle}{\emph{Proceedings of the ACM/IEEE 42nd international conference on software engineering}}. \bibinfo{pages}{481--492}.
\newblock


\bibitem[Fard and Mesbah(2013)]%
        {feedex}
\bibfield{author}{\bibinfo{person}{Amin~Milani Fard} {and} \bibinfo{person}{Ali Mesbah}.} \bibinfo{year}{2013}\natexlab{}.
\newblock \showarticletitle{Feedback-directed exploration of web applications to derive test models.}. In \bibinfo{booktitle}{\emph{ISSRE}}, Vol.~\bibinfo{volume}{13}. \bibinfo{pages}{278--287}.
\newblock


\bibitem[Garc{\'\i}a et~al\mbox{.}(2024)]%
        {garcia2024use}
\bibfield{author}{\bibinfo{person}{Boni Garc{\'\i}a}, \bibinfo{person}{Maurizio Leotta}, \bibinfo{person}{Filippo Ricca}, {and} \bibinfo{person}{Jim Whitehead}.} \bibinfo{year}{2024}\natexlab{}.
\newblock \showarticletitle{Use of chatgpt as an assistant in the end-to-end test script generation for android apps}. In \bibinfo{booktitle}{\emph{Proceedings of the 15th ACM International Workshop on Automating Test Case Design, Selection and Evaluation}}. \bibinfo{pages}{5--11}.
\newblock


\bibitem[Gou et~al\mbox{.}(2025)]%
        {uground}
\bibfield{author}{\bibinfo{person}{Boyu Gou}, \bibinfo{person}{Ruohan Wang}, \bibinfo{person}{Boyuan Zheng}, \bibinfo{person}{Yanan Xie}, \bibinfo{person}{Cheng Chang}, \bibinfo{person}{Yiheng Shu}, \bibinfo{person}{Huan Sun}, {and} \bibinfo{person}{Yu Su}.} \bibinfo{year}{2025}\natexlab{}.
\newblock \showarticletitle{Navigating the Digital World as Humans Do: Universal Visual Grounding for {GUI} Agents}. In \bibinfo{booktitle}{\emph{The Thirteenth International Conference on Learning Representations}}.
\newblock
\urldef\tempurl%
\url{https://openreview.net/forum?id=kxnoqaisCT}
\showURL{%
\tempurl}


\bibitem[gremlin.js(2014)]%
        {gremlin.js}
gremlin.js \bibinfo{year}{2014}\natexlab{}.
\newblock \bibinfo{howpublished}{\url{https://github.com/marmelab/gremlins.js/}}.
\newblock


\bibitem[Gu et~al\mbox{.}(2025a)]%
        {webrled}
\bibfield{author}{\bibinfo{person}{Zhiyu Gu}, \bibinfo{person}{Chenxu Liu}, \bibinfo{person}{Guoquan Wu}, \bibinfo{person}{Yifei Zhang}, \bibinfo{person}{ChenXi Yang}, \bibinfo{person}{Zheheng Liang}, \bibinfo{person}{Wei Chen}, {and} \bibinfo{person}{Jun Wei}.} \bibinfo{year}{2025}\natexlab{a}.
\newblock \showarticletitle{Deep Reinforcement Learning for Automated Web GUI Testing}.
\newblock \bibinfo{journal}{\emph{arXiv preprint arXiv:2504.19237}} (\bibinfo{year}{2025}).
\newblock


\bibitem[Gu et~al\mbox{.}(2025b)]%
        {uivenus}
\bibfield{author}{\bibinfo{person}{Zhangxuan Gu}, \bibinfo{person}{Zhengwen Zeng}, \bibinfo{person}{Zhenyu Xu}, \bibinfo{person}{Xingran Zhou}, \bibinfo{person}{Shuheng Shen}, \bibinfo{person}{Yunfei Liu}, \bibinfo{person}{Beitong Zhou}, \bibinfo{person}{Changhua Meng}, \bibinfo{person}{Tianyu Xia}, \bibinfo{person}{Weizhi Chen}, {et~al\mbox{.}}} \bibinfo{year}{2025}\natexlab{b}.
\newblock \showarticletitle{Ui-venus technical report: Building high-performance ui agents with rft}.
\newblock \bibinfo{journal}{\emph{arXiv preprint arXiv:2508.10833}} (\bibinfo{year}{2025}).
\newblock


\bibitem[https://www.qt.io/quality-assurance/squish(2003)]%
        {squish}
https://www.qt.io/quality-assurance/squish \bibinfo{year}{2003}\natexlab{}.
\newblock
\newblock
\urldef\tempurl%
\url{https://www.qt.io/quality-assurance/squish}
\showURL{%
\tempurl}


\bibitem[Hu et~al\mbox{.}(2018)]%
        {appflow}
\bibfield{author}{\bibinfo{person}{Gang Hu}, \bibinfo{person}{Linjie Zhu}, {and} \bibinfo{person}{Junfeng Yang}.} \bibinfo{year}{2018}\natexlab{}.
\newblock \showarticletitle{AppFlow: using machine learning to synthesize robust, reusable UI tests}. In \bibinfo{booktitle}{\emph{Proceedings of the 2018 26th ACM Joint Meeting on European Software Engineering Conference and Symposium on the Foundations of Software Engineering}}. \bibinfo{pages}{269--282}.
\newblock


\bibitem[Hu et~al\mbox{.}(2024)]%
        {AUITestAgent}
\bibfield{author}{\bibinfo{person}{Yongxiang Hu}, \bibinfo{person}{Xuan Wang}, \bibinfo{person}{Yingchuan Wang}, \bibinfo{person}{Yu Zhang}, \bibinfo{person}{Shiyu Guo}, \bibinfo{person}{Chaoyi Chen}, \bibinfo{person}{Xin Wang}, {and} \bibinfo{person}{Yangfan Zhou}.} \bibinfo{year}{2024}\natexlab{}.
\newblock \showarticletitle{Auitestagent: Automatic requirements oriented gui function testing}.
\newblock \bibinfo{journal}{\emph{arXiv preprint arXiv:2407.09018}} (\bibinfo{year}{2024}).
\newblock


\bibitem[Indico(2004)]%
        {indico}
Indico \bibinfo{year}{2004}\natexlab{}.
\newblock \bibinfo{howpublished}{\url{https://github.com/indico/indico}}.
\newblock


\bibitem[Invoice Ninja(2018)]%
        {invoiceninja}
Invoice Ninja \bibinfo{year}{2018}\natexlab{}.
\newblock \bibinfo{howpublished}{\url{https://github.com/invoiceninja/invoiceninja}}.
\newblock


\bibitem[LaVague(2024)]%
        {lavague}
LaVague \bibinfo{year}{2024}\natexlab{}.
\newblock \bibinfo{howpublished}{\url{https://github.com/lavague-ai/LaVague}}.
\newblock


\bibitem[Leotta et~al\mbox{.}(2024)]%
        {leotta2024ai}
\bibfield{author}{\bibinfo{person}{Maurizio Leotta}, \bibinfo{person}{Hafiz~Zeeshan Yousaf}, \bibinfo{person}{Filippo Ricca}, {and} \bibinfo{person}{Boni Garcia}.} \bibinfo{year}{2024}\natexlab{}.
\newblock \showarticletitle{Ai-generated test scripts for web e2e testing with chatgpt and copilot: A preliminary study}. In \bibinfo{booktitle}{\emph{Proceedings of the 28th International Conference on Evaluation and Assessment in Software Engineering}}. \bibinfo{pages}{339--344}.
\newblock


\bibitem[Li et~al\mbox{.}(2025b)]%
        {formfactory}
\bibfield{author}{\bibinfo{person}{Bobo Li}, \bibinfo{person}{Yuheng Wang}, \bibinfo{person}{Hao Fei}, \bibinfo{person}{Juncheng Li}, \bibinfo{person}{Wei Ji}, \bibinfo{person}{Mong-Li Lee}, {and} \bibinfo{person}{Wynne Hsu}.} \bibinfo{year}{2025}\natexlab{b}.
\newblock \showarticletitle{FormFactory: An Interactive Benchmarking Suite for Multimodal Form-Filling Agents}.
\newblock \bibinfo{journal}{\emph{arXiv preprint arXiv:2506.01520}} (\bibinfo{year}{2025}).
\newblock


\bibitem[Li et~al\mbox{.}(2025a)]%
        {screenseeker}
\bibfield{author}{\bibinfo{person}{Kaixin Li}, \bibinfo{person}{Ziyang Meng}, \bibinfo{person}{Hongzhan Lin}, \bibinfo{person}{Ziyang Luo}, \bibinfo{person}{Yuchen Tian}, \bibinfo{person}{Jing Ma}, \bibinfo{person}{Zhiyong Huang}, {and} \bibinfo{person}{Tat-Seng Chua}.} \bibinfo{year}{2025}\natexlab{a}.
\newblock \showarticletitle{Screenspot-pro: Gui grounding for professional high-resolution computer use}. In \bibinfo{booktitle}{\emph{Proceedings of the 33rd ACM International Conference on Multimedia}}. \bibinfo{pages}{8778--8786}.
\newblock


\bibitem[Liu et~al\mbox{.}(2025a)]%
        {Temac}
\bibfield{author}{\bibinfo{person}{Chenxu Liu}, \bibinfo{person}{Zhiyu Gu}, \bibinfo{person}{Guoquan Wu}, \bibinfo{person}{Ying Zhang}, \bibinfo{person}{Jun Wei}, {and} \bibinfo{person}{Tao Xie}.} \bibinfo{year}{2025}\natexlab{a}.
\newblock \showarticletitle{Temac: Multi-Agent Collaboration for Automated Web GUI Testing}.
\newblock \bibinfo{journal}{\emph{arXiv preprint arXiv:2506.00520}} (\bibinfo{year}{2025}).
\newblock


\bibitem[Liu et~al\mbox{.}(2025c)]%
        {judge}
\bibfield{author}{\bibinfo{person}{Chenxu Liu}, \bibinfo{person}{Junheng Wang}, \bibinfo{person}{Wei Yang}, \bibinfo{person}{Ying Zhang}, {and} \bibinfo{person}{Tao Xie}.} \bibinfo{year}{2025}\natexlab{c}.
\newblock \showarticletitle{Judge: Effective State Abstraction for Guiding Automated Web GUI Testing}.
\newblock \bibinfo{journal}{\emph{ACM Transactions on Software Engineering and Methodology}} (\bibinfo{year}{2025}).
\newblock


\bibitem[Liu et~al\mbox{.}(2025b)]%
        {guipilot}
\bibfield{author}{\bibinfo{person}{Ruofan Liu}, \bibinfo{person}{Xiwen Teoh}, \bibinfo{person}{Yun Lin}, \bibinfo{person}{Guanjie Chen}, \bibinfo{person}{Ruofei Ren}, \bibinfo{person}{Denys Poshyvanyk}, {and} \bibinfo{person}{Jin~Song Dong}.} \bibinfo{year}{2025}\natexlab{b}.
\newblock \showarticletitle{GUIPilot: A Consistency-Based Mobile GUI Testing Approach for Detecting Application-Specific Bugs}.
\newblock \bibinfo{journal}{\emph{Proceedings of the ACM on Software Engineering}} \bibinfo{volume}{2}, \bibinfo{number}{ISSTA} (\bibinfo{year}{2025}), \bibinfo{pages}{753--776}.
\newblock


\bibitem[Liu et~al\mbox{.}(2024d)]%
        {liu2024wefix}
\bibfield{author}{\bibinfo{person}{Xinyue Liu}, \bibinfo{person}{Zihe Song}, \bibinfo{person}{Weike Fang}, \bibinfo{person}{Wei Yang}, {and} \bibinfo{person}{Weihang Wang}.} \bibinfo{year}{2024}\natexlab{d}.
\newblock \showarticletitle{Wefix: Intelligent automatic generation of explicit waits for efficient web end-to-end flaky tests}. In \bibinfo{booktitle}{\emph{Proceedings of the ACM Web Conference 2024}}. \bibinfo{pages}{3043--3052}.
\newblock


\bibitem[Liu et~al\mbox{.}(2023a)]%
        {QTypist}
\bibfield{author}{\bibinfo{person}{Zhe Liu}, \bibinfo{person}{Chunyang Chen}, \bibinfo{person}{Junjie Wang}, \bibinfo{person}{Xing Che}, \bibinfo{person}{Yuekai Huang}, \bibinfo{person}{Jun Hu}, {and} \bibinfo{person}{Qing Wang}.} \bibinfo{year}{2023}\natexlab{a}.
\newblock \showarticletitle{Fill in the blank: Context-aware automated text input generation for mobile gui testing}. In \bibinfo{booktitle}{\emph{2023 IEEE/ACM 45th International Conference on Software Engineering (ICSE)}}. IEEE, \bibinfo{pages}{1355--1367}.
\newblock


\bibitem[Liu et~al\mbox{.}(2023b)]%
        {GPTDroid}
\bibfield{author}{\bibinfo{person}{Zhe Liu}, \bibinfo{person}{Chunyang Chen}, \bibinfo{person}{Junjie Wang}, \bibinfo{person}{Mengzhuo Chen}, \bibinfo{person}{Boyu Wu}, \bibinfo{person}{Xing Che}, \bibinfo{person}{Dandan Wang}, {and} \bibinfo{person}{Qing Wang}.} \bibinfo{year}{2023}\natexlab{b}.
\newblock \showarticletitle{Chatting with gpt-3 for zero-shot human-like mobile automated gui testing}.
\newblock \bibinfo{journal}{\emph{arXiv preprint arXiv:2305.09434}} (\bibinfo{year}{2023}).
\newblock


\bibitem[Liu et~al\mbox{.}(2024a)]%
        {GPTDroid-v2}
\bibfield{author}{\bibinfo{person}{Zhe Liu}, \bibinfo{person}{Chunyang Chen}, \bibinfo{person}{Junjie Wang}, \bibinfo{person}{Mengzhuo Chen}, \bibinfo{person}{Boyu Wu}, \bibinfo{person}{Xing Che}, \bibinfo{person}{Dandan Wang}, {and} \bibinfo{person}{Qing Wang}.} \bibinfo{year}{2024}\natexlab{a}.
\newblock \showarticletitle{Make llm a testing expert: Bringing human-like interaction to mobile gui testing via functionality-aware decisions}. In \bibinfo{booktitle}{\emph{Proceedings of the IEEE/ACM 46th International Conference on Software Engineering}}. \bibinfo{pages}{1--13}.
\newblock


\bibitem[Liu et~al\mbox{.}(2024b)]%
        {InputBlaster}
\bibfield{author}{\bibinfo{person}{Zhe Liu}, \bibinfo{person}{Chunyang Chen}, \bibinfo{person}{Junjie Wang}, \bibinfo{person}{Mengzhuo Chen}, \bibinfo{person}{Boyu Wu}, \bibinfo{person}{Zhilin Tian}, \bibinfo{person}{Yuekai Huang}, \bibinfo{person}{Jun Hu}, {and} \bibinfo{person}{Qing Wang}.} \bibinfo{year}{2024}\natexlab{b}.
\newblock \showarticletitle{Testing the limits: Unusual text inputs generation for mobile app crash detection with large language model}. In \bibinfo{booktitle}{\emph{Proceedings of the IEEE/ACM 46th International conference on software engineering}}. \bibinfo{pages}{1--12}.
\newblock


\bibitem[Liu et~al\mbox{.}(2024c)]%
        {Trident}
\bibfield{author}{\bibinfo{person}{Zhe Liu}, \bibinfo{person}{Cheng Li}, \bibinfo{person}{Chunyang Chen}, \bibinfo{person}{Junjie Wang}, \bibinfo{person}{Mengzhuo Chen}, \bibinfo{person}{Boyu Wu}, \bibinfo{person}{Yawen Wang}, \bibinfo{person}{Jun Hu}, {and} \bibinfo{person}{Qing Wang}.} \bibinfo{year}{2024}\natexlab{c}.
\newblock \showarticletitle{Seeing is Believing: Vision-driven Non-crash Functional Bug Detection for Mobile Apps}.
\newblock \bibinfo{journal}{\emph{arXiv preprint arXiv:2407.03037}} (\bibinfo{year}{2024}).
\newblock


\bibitem[Lu et~al\mbox{.}(2024)]%
        {omniparser}
\bibfield{author}{\bibinfo{person}{Yadong Lu}, \bibinfo{person}{Jianwei Yang}, \bibinfo{person}{Yelong Shen}, {and} \bibinfo{person}{Ahmed Awadallah}.} \bibinfo{year}{2024}\natexlab{}.
\newblock \showarticletitle{Omniparser for pure vision based gui agent}.
\newblock \bibinfo{journal}{\emph{arXiv preprint arXiv:2408.00203}} (\bibinfo{year}{2024}).
\newblock


\bibitem[Lu et~al\mbox{.}(2025)]%
        {UXAgent}
\bibfield{author}{\bibinfo{person}{Yuxuan Lu}, \bibinfo{person}{Bingsheng Yao}, \bibinfo{person}{Hansu Gu}, \bibinfo{person}{Jing Huang}, \bibinfo{person}{Zheshen~Jessie Wang}, \bibinfo{person}{Yang Li}, \bibinfo{person}{Jiri Gesi}, \bibinfo{person}{Qi He}, \bibinfo{person}{Toby Jia-Jun Li}, {and} \bibinfo{person}{Dakuo Wang}.} \bibinfo{year}{2025}\natexlab{}.
\newblock \showarticletitle{Uxagent: An llm agent-based usability testing framework for web design}. In \bibinfo{booktitle}{\emph{Proceedings of the Extended Abstracts of the CHI Conference on Human Factors in Computing Systems}}. \bibinfo{pages}{1--12}.
\newblock


\bibitem[Mao et~al\mbox{.}(2016)]%
        {sapienz}
\bibfield{author}{\bibinfo{person}{Ke Mao}, \bibinfo{person}{Mark Harman}, {and} \bibinfo{person}{Yue Jia}.} \bibinfo{year}{2016}\natexlab{}.
\newblock \showarticletitle{Sapienz: Multi-objective automated testing for android applications}. In \bibinfo{booktitle}{\emph{Proceedings of the 25th international symposium on software testing and analysis}}. \bibinfo{pages}{94--105}.
\newblock


\bibitem[Mariani et~al\mbox{.}(2011)]%
        {autoblacktest}
\bibfield{author}{\bibinfo{person}{Leonardo Mariani}, \bibinfo{person}{Mauro Pezz{\`e}}, \bibinfo{person}{Oliviero Riganelli}, {and} \bibinfo{person}{Mauro Santoro}.} \bibinfo{year}{2011}\natexlab{}.
\newblock \showarticletitle{AutoBlackTest: a tool for automatic black-box testing}. In \bibinfo{booktitle}{\emph{Proceedings of the 33rd international conference on software engineering}}. \bibinfo{pages}{1013--1015}.
\newblock


\bibitem[Mesbah et~al\mbox{.}(2008)]%
        {crawljax}
\bibfield{author}{\bibinfo{person}{Ali Mesbah}, \bibinfo{person}{Engin Bozdag}, {and} \bibinfo{person}{Arie Van~Deursen}.} \bibinfo{year}{2008}\natexlab{}.
\newblock \showarticletitle{Crawling Ajax by inferring user interface state changes}. In \bibinfo{booktitle}{\emph{2008 eighth international conference on web engineering}}. IEEE, \bibinfo{pages}{122--134}.
\newblock


\bibitem[Mesbah et~al\mbox{.}(2011)]%
        {atusa}
\bibfield{author}{\bibinfo{person}{Ali Mesbah}, \bibinfo{person}{Arie Van~Deursen}, {and} \bibinfo{person}{Danny Roest}.} \bibinfo{year}{2011}\natexlab{}.
\newblock \showarticletitle{Invariant-based automatic testing of modern web applications}.
\newblock \bibinfo{journal}{\emph{IEEE Transactions on Software Engineering}} \bibinfo{volume}{38}, \bibinfo{number}{1} (\bibinfo{year}{2011}), \bibinfo{pages}{35--53}.
\newblock


\bibitem[Monkey(2023)]%
        {Monkey}
Monkey \bibinfo{year}{2023}\natexlab{}.
\newblock \bibinfo{howpublished}{\url{https://developer.android.com/studio/test/other-testing-tools/monkey}}.
\newblock


\bibitem[Olianas et~al\mbox{.}(2021)]%
        {stile}
\bibfield{author}{\bibinfo{person}{Dario Olianas}, \bibinfo{person}{Maurizio Leotta}, \bibinfo{person}{Filippo Ricca}, \bibinfo{person}{Matteo Biagiola}, {and} \bibinfo{person}{Paolo Tonella}.} \bibinfo{year}{2021}\natexlab{}.
\newblock \showarticletitle{STILE: a tool for parallel execution of E2E web test scripts}. In \bibinfo{booktitle}{\emph{2021 14th IEEE Conference on Software Testing, Verification and Validation (ICST)}}. IEEE, \bibinfo{pages}{460--465}.
\newblock


\bibitem[Pei et~al\mbox{.}(2025)]%
        {pei2025non}
\bibfield{author}{\bibinfo{person}{Yu Pei}, \bibinfo{person}{Jeongju Sohn}, \bibinfo{person}{Sarra Habchi}, {and} \bibinfo{person}{Mike Papadakis}.} \bibinfo{year}{2025}\natexlab{}.
\newblock \showarticletitle{Non-flaky and nearly optimal time-based treatment of asynchronous wait web tests}.
\newblock \bibinfo{journal}{\emph{ACM Transactions on Software Engineering and Methodology}} \bibinfo{volume}{34}, \bibinfo{number}{2} (\bibinfo{year}{2025}), \bibinfo{pages}{1--29}.
\newblock


\bibitem[Peldszus et~al\mbox{.}(2023)]%
        {robotbt}
\bibfield{author}{\bibinfo{person}{Sven Peldszus}, \bibinfo{person}{Noubar Akopian}, {and} \bibinfo{person}{Thorsten Berger}.} \bibinfo{year}{2023}\natexlab{}.
\newblock \showarticletitle{RobotBT: Behavior-tree-based test-case specification for the robot framework}. In \bibinfo{booktitle}{\emph{Proceedings of the 32nd ACM SIGSOFT International Symposium on Software Testing and Analysis}}. \bibinfo{pages}{1503--1506}.
\newblock


\bibitem[Peng et~al\mbox{.}(2024)]%
        {hawkeye}
\bibfield{author}{\bibinfo{person}{Chao Peng}, \bibinfo{person}{Zhengwei Lv}, \bibinfo{person}{Jiarong Fu}, \bibinfo{person}{Jiayuan Liang}, \bibinfo{person}{Zhao Zhang}, \bibinfo{person}{Ajitha Rajan}, {and} \bibinfo{person}{Ping Yang}.} \bibinfo{year}{2024}\natexlab{}.
\newblock \showarticletitle{Hawkeye: Change-targeted testing for android apps based on deep reinforcement learning}. In \bibinfo{booktitle}{\emph{Proceedings of the 46th International Conference on Software Engineering: Software Engineering in Practice}}. \bibinfo{pages}{298--308}.
\newblock


\bibitem[Prestashop(2007)]%
        {Prestashop}
Prestashop \bibinfo{year}{2007}\natexlab{}.
\newblock \bibinfo{howpublished}{\url{https://github.com/saleor/saleor}}.
\newblock


\bibitem[Progressive Web Apps Market Size, Share \& Trends Analysis Report, 2024–2030(2024)]%
        {grandview2024pwa}
Progressive Web Apps Market Size, Share \& Trends Analysis Report, 2024–2030 \bibinfo{year}{2024}\natexlab{}.
\newblock
\newblock
\urldef\tempurl%
\url{https://www.grandviewresearch.com/industry-analysis/progressive-web-apps-pwa-market-report}
\showURL{%
\tempurl}


\bibitem[Ran et~al\mbox{.}(2024)]%
        {Guardian}
\bibfield{author}{\bibinfo{person}{Dezhi Ran}, \bibinfo{person}{Hao Wang}, \bibinfo{person}{Zihe Song}, \bibinfo{person}{Mengzhou Wu}, \bibinfo{person}{Yuan Cao}, \bibinfo{person}{Ying Zhang}, \bibinfo{person}{Wei Yang}, {and} \bibinfo{person}{Tao Xie}.} \bibinfo{year}{2024}\natexlab{}.
\newblock \showarticletitle{Guardian: A runtime framework for LLM-based UI exploration}. In \bibinfo{booktitle}{\emph{Proceedings of the 33rd ACM SIGSOFT International Symposium on Software Testing and Analysis}}. \bibinfo{pages}{958--970}.
\newblock


\bibitem[RSpec(2007)]%
        {rspec}
RSpec \bibinfo{year}{2007}\natexlab{}.
\newblock
\newblock
\urldef\tempurl%
\url{https://rspec.info/}
\showURL{%
\tempurl}


\bibitem[Salma et~al\mbox{.}(2024)]%
        {guievo}
\bibfield{author}{\bibinfo{person}{Sabiha Salma}, \bibinfo{person}{SM~Hasan Mansur}, \bibinfo{person}{Yule Zhang}, {and} \bibinfo{person}{Kevin Moran}.} \bibinfo{year}{2024}\natexlab{}.
\newblock \showarticletitle{GuiEvo: Automated Evolution of Mobile App UIs}. In \bibinfo{booktitle}{\emph{Proceedings of the 21st International Conference on Mining Software Repositories}}. \bibinfo{pages}{335--347}.
\newblock


\bibitem[Shahbandeh et~al\mbox{.}(2024)]%
        {naviqate}
\bibfield{author}{\bibinfo{person}{Mobina Shahbandeh}, \bibinfo{person}{Parsa Alian}, \bibinfo{person}{Noor Nashid}, {and} \bibinfo{person}{Ali Mesbah}.} \bibinfo{year}{2024}\natexlab{}.
\newblock \showarticletitle{Naviqate: Functionality-guided web application navigation}.
\newblock \bibinfo{journal}{\emph{arXiv preprint arXiv:2409.10741}} (\bibinfo{year}{2024}).
\newblock


\bibitem[Shao et~al\mbox{.}(2021)]%
        {webevo}
\bibfield{author}{\bibinfo{person}{Fei Shao}, \bibinfo{person}{Rui Xu}, \bibinfo{person}{Wasif Haque}, \bibinfo{person}{Jingwei Xu}, \bibinfo{person}{Ying Zhang}, \bibinfo{person}{Wei Yang}, \bibinfo{person}{Yanfang Ye}, {and} \bibinfo{person}{Xusheng Xiao}.} \bibinfo{year}{2021}\natexlab{}.
\newblock \showarticletitle{Webevo: taming web application evolution via detecting semantic structure changes}. In \bibinfo{booktitle}{\emph{Proceedings of the 30th ACM SIGSOFT International Symposium on Software Testing and Analysis}}. \bibinfo{pages}{16--28}.
\newblock


\bibitem[Sherin et~al\mbox{.}(2023)]%
        {qexplore}
\bibfield{author}{\bibinfo{person}{Salman Sherin}, \bibinfo{person}{Asmar Muqeet}, \bibinfo{person}{Muhammad~Uzair Khan}, {and} \bibinfo{person}{Muhammad~Zohaib Iqbal}.} \bibinfo{year}{2023}\natexlab{}.
\newblock \showarticletitle{QExplore: An exploration strategy for dynamic web applications using guided search}.
\newblock \bibinfo{journal}{\emph{Journal of Systems and Software}}  \bibinfo{volume}{195} (\bibinfo{year}{2023}), \bibinfo{pages}{111512}.
\newblock


\bibitem[State of Software Quality Report(2024)]%
        {SOSQR-2024}
State of Software Quality Report \bibinfo{year}{2024}\natexlab{}.
\newblock
\newblock
\urldef\tempurl%
\url{https://katalon.com/reports/state-quality-2024}
\showURL{%
\tempurl}


\bibitem[Stocco et~al\mbox{.}(2023)]%
        {webembed}
\bibfield{author}{\bibinfo{person}{Andrea Stocco}, \bibinfo{person}{Alexandra Willi}, \bibinfo{person}{Luigi Libero~Lucio Starace}, \bibinfo{person}{Matteo Biagiola}, {and} \bibinfo{person}{Paolo Tonella}.} \bibinfo{year}{2023}\natexlab{}.
\newblock \showarticletitle{Neural embeddings for web testing}.
\newblock \bibinfo{journal}{\emph{arXiv preprint arXiv:2306.07400}} (\bibinfo{year}{2023}).
\newblock


\bibitem[Su et~al\mbox{.}(2020)]%
        {droiddefects}
\bibfield{author}{\bibinfo{person}{Ting Su}, \bibinfo{person}{Lingling Fan}, \bibinfo{person}{Sen Chen}, \bibinfo{person}{Yang Liu}, \bibinfo{person}{Lihua Xu}, \bibinfo{person}{Geguang Pu}, {and} \bibinfo{person}{Zhendong Su}.} \bibinfo{year}{2020}\natexlab{}.
\newblock \showarticletitle{Why my app crashes? understanding and benchmarking framework-specific exceptions of android apps}.
\newblock \bibinfo{journal}{\emph{IEEE Transactions on Software Engineering}} \bibinfo{volume}{48}, \bibinfo{number}{4} (\bibinfo{year}{2020}), \bibinfo{pages}{1115--1137}.
\newblock


\bibitem[Su et~al\mbox{.}(2017)]%
        {stoat}
\bibfield{author}{\bibinfo{person}{Ting Su}, \bibinfo{person}{Guozhu Meng}, \bibinfo{person}{Yuting Chen}, \bibinfo{person}{Ke Wu}, \bibinfo{person}{Weiming Yang}, \bibinfo{person}{Yao Yao}, \bibinfo{person}{Geguang Pu}, \bibinfo{person}{Yang Liu}, {and} \bibinfo{person}{Zhendong Su}.} \bibinfo{year}{2017}\natexlab{}.
\newblock \showarticletitle{Guided, stochastic model-based GUI testing of Android apps}. In \bibinfo{booktitle}{\emph{Proceedings of the 2017 11th joint meeting on foundations of software engineering}}. \bibinfo{pages}{245--256}.
\newblock


\bibitem[Su et~al\mbox{.}(2021)]%
        {themis}
\bibfield{author}{\bibinfo{person}{Ting Su}, \bibinfo{person}{Jue Wang}, {and} \bibinfo{person}{Zhendong Su}.} \bibinfo{year}{2021}\natexlab{}.
\newblock \showarticletitle{Benchmarking automated gui testing for android against real-world bugs}. In \bibinfo{booktitle}{\emph{Proceedings of the 29th ACM Joint Meeting on European Software Engineering Conference and Symposium on the Foundations of Software Engineering}}. \bibinfo{pages}{119--130}.
\newblock


\bibitem[Taeb et~al\mbox{.}(2024)]%
        {Axnav}
\bibfield{author}{\bibinfo{person}{Maryam Taeb}, \bibinfo{person}{Amanda Swearngin}, \bibinfo{person}{Eldon Schoop}, \bibinfo{person}{Ruijia Cheng}, \bibinfo{person}{Yue Jiang}, {and} \bibinfo{person}{Jeffrey Nichols}.} \bibinfo{year}{2024}\natexlab{}.
\newblock \showarticletitle{Axnav: Replaying accessibility tests from natural language}. In \bibinfo{booktitle}{\emph{Proceedings of the 2024 CHI Conference on Human Factors in Computing Systems}}. \bibinfo{pages}{1--16}.
\newblock


\bibitem[The Failed Launch Of www.HealthCare.gov(2016)]%
        {healthcaregov-fail}
The Failed Launch Of www.HealthCare.gov \bibinfo{year}{2016}\natexlab{}.
\newblock
\newblock
\urldef\tempurl%
\url{https://d3.harvard.edu/platform-rctom/submission/the-failed-launch-of-www-healthcare-gov/}
\showURL{%
\tempurl}


\bibitem[The Payroll System That Cost Queensland Health AU1.25 Billion({[n.\,d.]})]%
        {queenslandpayroll-fail}
The Payroll System That Cost Queensland Health AU1.25 Billion \bibinfo{year}{[n.\,d.]}\natexlab{}.
\newblock
\newblock
\urldef\tempurl%
\url{https://www.henricodolfing.com/2019/12/project-failure-case-study-queensland-health.html}
\showURL{%
\tempurl}


\bibitem[Wang et~al\mbox{.}(2024a)]%
        {wang2024leveraging}
\bibfield{author}{\bibinfo{person}{Siyi Wang}, \bibinfo{person}{Sinan Wang}, \bibinfo{person}{Yujia Fan}, \bibinfo{person}{Xiaolei Li}, {and} \bibinfo{person}{Yepang Liu}.} \bibinfo{year}{2024}\natexlab{a}.
\newblock \showarticletitle{Leveraging large vision-language model for better automatic web GUI testing}. In \bibinfo{booktitle}{\emph{2024 IEEE International Conference on Software Maintenance and Evolution (ICSME)}}. IEEE, \bibinfo{pages}{125--137}.
\newblock


\bibitem[Wang et~al\mbox{.}(2025)]%
        {wang2025mmbench}
\bibfield{author}{\bibinfo{person}{Xuehui Wang}, \bibinfo{person}{Zhenyu Wu}, \bibinfo{person}{JingJing Xie}, \bibinfo{person}{Zichen Ding}, \bibinfo{person}{Bowen Yang}, \bibinfo{person}{Zehao Li}, \bibinfo{person}{Zhaoyang Liu}, \bibinfo{person}{Qingyun Li}, \bibinfo{person}{Xuan Dong}, \bibinfo{person}{Zhe Chen}, {et~al\mbox{.}}} \bibinfo{year}{2025}\natexlab{}.
\newblock \showarticletitle{MMBench-GUI: Hierarchical Multi-Platform Evaluation Framework for GUI Agents}.
\newblock \bibinfo{journal}{\emph{arXiv preprint arXiv:2507.19478}} (\bibinfo{year}{2025}).
\newblock


\bibitem[Wang et~al\mbox{.}(2024b)]%
        {xuat-copilot}
\bibfield{author}{\bibinfo{person}{Zhitao Wang}, \bibinfo{person}{Wei Wang}, \bibinfo{person}{Zirao Li}, \bibinfo{person}{Long Wang}, \bibinfo{person}{Can Yi}, \bibinfo{person}{Xinjie Xu}, \bibinfo{person}{Luyang Cao}, \bibinfo{person}{Hanjing Su}, \bibinfo{person}{Shouzhi Chen}, {and} \bibinfo{person}{Jun Zhou}.} \bibinfo{year}{2024}\natexlab{b}.
\newblock \showarticletitle{Xuat-copilot: Multi-agent collaborative system for automated user acceptance testing with large language model}.
\newblock \bibinfo{journal}{\emph{arXiv preprint arXiv:2401.02705}} (\bibinfo{year}{2024}).
\newblock


\bibitem[White et~al\mbox{.}(2019)]%
        {white2019improving}
\bibfield{author}{\bibinfo{person}{Thomas~D White}, \bibinfo{person}{Gordon Fraser}, {and} \bibinfo{person}{Guy~J Brown}.} \bibinfo{year}{2019}\natexlab{}.
\newblock \showarticletitle{Improving random GUI testing with image-based widget detection}. In \bibinfo{booktitle}{\emph{Proceedings of the 28th ACM SIGSOFT international symposium on software testing and analysis}}. \bibinfo{pages}{307--317}.
\newblock


\bibitem[Xiong et~al\mbox{.}(2024)]%
        {kea}
\bibfield{author}{\bibinfo{person}{Yiheng Xiong}, \bibinfo{person}{Ting Su}, \bibinfo{person}{Jue Wang}, \bibinfo{person}{Jingling Sun}, \bibinfo{person}{Geguang Pu}, {and} \bibinfo{person}{Zhendong Su}.} \bibinfo{year}{2024}\natexlab{}.
\newblock \showarticletitle{General and practical property-based testing for android apps}. In \bibinfo{booktitle}{\emph{Proceedings of the 39th IEEE/ACM International Conference on Automated Software Engineering}}. \bibinfo{pages}{53--64}.
\newblock


\bibitem[Yandrapally et~al\mbox{.}(2020)]%
        {ndstudy}
\bibfield{author}{\bibinfo{person}{Rahulkrishna Yandrapally}, \bibinfo{person}{Andrea Stocco}, {and} \bibinfo{person}{Ali Mesbah}.} \bibinfo{year}{2020}\natexlab{}.
\newblock \showarticletitle{Near-duplicate detection in web app model inference}. In \bibinfo{booktitle}{\emph{Proceedings of the ACM/IEEE 42nd international conference on software engineering}}. \bibinfo{pages}{186--197}.
\newblock


\bibitem[Yandrapally and Mesbah(2022)]%
        {fraggen}
\bibfield{author}{\bibinfo{person}{Rahul~Krishna Yandrapally} {and} \bibinfo{person}{Ali Mesbah}.} \bibinfo{year}{2022}\natexlab{}.
\newblock \showarticletitle{Fragment-based test generation for web apps}.
\newblock \bibinfo{journal}{\emph{IEEE Transactions on Software Engineering}} \bibinfo{volume}{49}, \bibinfo{number}{3} (\bibinfo{year}{2022}), \bibinfo{pages}{1086--1101}.
\newblock


\bibitem[Yoon et~al\mbox{.}(2024)]%
        {DroidAgent}
\bibfield{author}{\bibinfo{person}{Juyeon Yoon}, \bibinfo{person}{Robert Feldt}, {and} \bibinfo{person}{Shin Yoo}.} \bibinfo{year}{2024}\natexlab{}.
\newblock \showarticletitle{Intent-driven mobile gui testing with autonomous large language model agents}. In \bibinfo{booktitle}{\emph{2024 IEEE Conference on Software Testing, Verification and Validation (ICST)}}. IEEE, \bibinfo{pages}{129--139}.
\newblock


\bibitem[Yu et~al\mbox{.}(2024a)]%
        {robotest}
\bibfield{author}{\bibinfo{person}{Shengcheng Yu}, \bibinfo{person}{Chunrong Fang}, \bibinfo{person}{Mingzhe Du}, \bibinfo{person}{Yuchen Ling}, \bibinfo{person}{Zhenyu Chen}, {and} \bibinfo{person}{Zhendong Su}.} \bibinfo{year}{2024}\natexlab{a}.
\newblock \showarticletitle{Practical non-intrusive GUI exploration testing with visual-based robotic arms}. In \bibinfo{booktitle}{\emph{Proceedings of the IEEE/ACM 46th International Conference on Software Engineering}}. \bibinfo{pages}{1--13}.
\newblock


\bibitem[Yu et~al\mbox{.}(2024b)]%
        {pirltest}
\bibfield{author}{\bibinfo{person}{Shengcheng Yu}, \bibinfo{person}{Chunrong Fang}, \bibinfo{person}{Xin Li}, \bibinfo{person}{Yuchen Ling}, \bibinfo{person}{Zhenyu Chen}, {and} \bibinfo{person}{Zhendong Su}.} \bibinfo{year}{2024}\natexlab{b}.
\newblock \showarticletitle{Effective, platform-independent gui testing via image embedding and reinforcement learning}.
\newblock \bibinfo{journal}{\emph{ACM Transactions on Software Engineering and Methodology}} \bibinfo{volume}{33}, \bibinfo{number}{7} (\bibinfo{year}{2024}), \bibinfo{pages}{1--27}.
\newblock


\bibitem[Yu et~al\mbox{.}(2022)]%
        {unirl}
\bibfield{author}{\bibinfo{person}{Shengcheng Yu}, \bibinfo{person}{Chunrong Fang}, \bibinfo{person}{Yulei Liu}, \bibinfo{person}{Ziqian Zhang}, \bibinfo{person}{Yexiao Yun}, \bibinfo{person}{Xin Li}, {and} \bibinfo{person}{Zhenyu Chen}.} \bibinfo{year}{2022}\natexlab{}.
\newblock \showarticletitle{Universally Adaptive Cross-Platform Reinforcement Learning Testing via GUI Image Understanding}.
\newblock \bibinfo{journal}{\emph{arXiv preprint arXiv:2208.09116}} (\bibinfo{year}{2022}).
\newblock


\bibitem[Zhang et~al\mbox{.}(2024)]%
        {zhang2024towards}
\bibfield{author}{\bibinfo{person}{Haonan Zhang}, \bibinfo{person}{Lizhi Liao}, \bibinfo{person}{Zishuo Ding}, \bibinfo{person}{Weiyi Shang}, \bibinfo{person}{Nidhi Narula}, \bibinfo{person}{Catalin Sporea}, \bibinfo{person}{Andrei Toma}, {and} \bibinfo{person}{Sarah Sajedi}.} \bibinfo{year}{2024}\natexlab{}.
\newblock \showarticletitle{Towards a Robust Waiting Strategy for Web GUI Testing for an Industrial Software System}. In \bibinfo{booktitle}{\emph{Proceedings of the 39th IEEE/ACM International Conference on Automated Software Engineering}}. \bibinfo{pages}{2065--2076}.
\newblock


\bibitem[Zhang et~al\mbox{.}(2025)]%
        {a11yscan}
\bibfield{author}{\bibinfo{person}{Yuxin Zhang}, \bibinfo{person}{Sen Chen}, \bibinfo{person}{Xiaofei Xie}, \bibinfo{person}{Zibo Liu}, {and} \bibinfo{person}{Lingling Fan}.} \bibinfo{year}{2025}\natexlab{}.
\newblock \showarticletitle{Scenario-Driven and Context-Aware Automated Accessibility Testing for Android Apps}. In \bibinfo{booktitle}{\emph{2025 IEEE/ACM 47th International Conference on Software Engineering (ICSE)}}. IEEE Computer Society, \bibinfo{pages}{630--630}.
\newblock


\bibitem[Zhao et~al\mbox{.}(2024)]%
        {guitestingarena}
\bibfield{author}{\bibinfo{person}{Kangjia Zhao}, \bibinfo{person}{Jiahui Song}, \bibinfo{person}{Leigang Sha}, \bibinfo{person}{Haozhan Shen}, \bibinfo{person}{Zhi Chen}, \bibinfo{person}{Tiancheng Zhao}, \bibinfo{person}{Xiubo Liang}, {and} \bibinfo{person}{Jianwei Yin}.} \bibinfo{year}{2024}\natexlab{}.
\newblock \showarticletitle{Gui testing arena: A unified benchmark for advancing autonomous gui testing agent}.
\newblock \bibinfo{journal}{\emph{arXiv preprint arXiv:2412.18426}} (\bibinfo{year}{2024}).
\newblock


\bibitem[Zhao et~al\mbox{.}(2025)]%
        {LLM-Explorer}
\bibfield{author}{\bibinfo{person}{Shanhui Zhao}, \bibinfo{person}{Hao Wen}, \bibinfo{person}{Wenjie Du}, \bibinfo{person}{Cheng Liang}, \bibinfo{person}{Yunxin Liu}, \bibinfo{person}{Xiaozhou Ye}, \bibinfo{person}{Ye Ouyang}, {and} \bibinfo{person}{Yuanchun Li}.} \bibinfo{year}{2025}\natexlab{}.
\newblock \showarticletitle{LLM-Explorer: Towards Efficient and Affordable LLM-based Exploration for Mobile Apps}.
\newblock \bibinfo{journal}{\emph{arXiv preprint arXiv:2505.10593}} (\bibinfo{year}{2025}).
\newblock


\bibitem[Zheng et~al\mbox{.}(2021)]%
        {webexplor}
\bibfield{author}{\bibinfo{person}{Yan Zheng}, \bibinfo{person}{Yi Liu}, \bibinfo{person}{Xiaofei Xie}, \bibinfo{person}{Yepang Liu}, \bibinfo{person}{Lei Ma}, \bibinfo{person}{Jianye Hao}, {and} \bibinfo{person}{Yang Liu}.} \bibinfo{year}{2021}\natexlab{}.
\newblock \showarticletitle{Automatic web testing using curiosity-driven reinforcement learning}. In \bibinfo{booktitle}{\emph{2021 IEEE/ACM 43rd International Conference on Software Engineering (ICSE)}}. IEEE, \bibinfo{pages}{423--435}.
\newblock


\bibitem[Zimmermann and Koziolek(2023)]%
        {zimmermann2023gui}
\bibfield{author}{\bibinfo{person}{Daniel Zimmermann} {and} \bibinfo{person}{Anne Koziolek}.} \bibinfo{year}{2023}\natexlab{}.
\newblock \showarticletitle{Gui-based software testing: An automated approach using gpt-4 and selenium webdriver}. In \bibinfo{booktitle}{\emph{2023 38th IEEE/ACM International Conference on Automated Software Engineering Workshops (ASEW)}}. IEEE, \bibinfo{pages}{171--174}.
\newblock


\end{thebibliography}

% \newpage
% \appendix

% \section{Detection of Real-world Bugs}
% \label{appendix:real_world_bugs}

% \input{tables/real_bugs}

\end{document}